\documentclass[longauth]{aa}  
\usepackage{graphicx}
\usepackage{txfonts}
\usepackage[colorlinks=true,citecolor=blue]{hyperref}

\usepackage{multirow}
\usepackage{xcolor}
\usepackage{caption}
\usepackage{subcaption}
\usepackage{threeparttable}
\usepackage{rotating}
\usepackage{rotfloat}
\usepackage{float}
\usepackage{moresize}
\usepackage{wrapfig}
\usepackage{multicol}
\usepackage{wrapfig}
\usepackage[normalem]{ulem}
\usepackage{graphicx}
\usepackage{txfonts}
\usepackage[T1]{fontenc}
 \usepackage{graphicx}  
\usepackage{amsmath}    
 \usepackage{color}
\usepackage{amssymb}
\usepackage{epstopdf}
\usepackage{fontawesome5}
\usepackage{longtable}
\usepackage{natbib}
\bibpunct{(}{)}{;}{a}{}{,} 
\usepackage{url}

\newcommand{\vsini}{$V \sin i$}

\newcommand{\teff}{$T_{\rm eff}$}
\newcommand{\logg}{log\,{\it g$_\star$}}

\newcommand{\feh}{[Fe/H]}

\newcommand{\kms}{km\,s$^{-1}$}
\newcommand{\ms}{m~s$^{-1}$}

\newcommand{\gc}{g~cm$^{-3}$}

\newcommand{\rstar}{$R_\star$}

\newcommand{\toioneb}{TOI-2420\,b}
\newcommand{\toitwob}{TOI-2485\,b}
\newcommand{\pyaneti}{\href{https://github.com/oscaribv/pyaneti}{\texttt{pyaneti}\,\faGithub}}

\begin{document}

\title{Mass determination of two Jupiter-sized planets orbiting slightly evolved stars: TOI-2420\,b and TOI-2485\,b }
\authorrunning{}


\author{Ilaria Carleo\inst{\ref{INAF-oato},\ref{iac},\ref{ull}},
          Oscar Barr{\'a}gan\inst{\ref{phy_uni_oxford}}, 
          Carina~M.~Persson\inst{\ref{OSO}},
          Malcolm Fridlund\inst{\ref{OSO},\ref{Leiden}},
          Kristine~W.\,F.~Lam\inst{\ref{DLR}},
          Sergio Messina\inst{\ref{INAF-Catania}},
          Davide Gandolfi\inst{\ref{UniTo}},
          Alexis M. S. Smith\inst{\ref{DLR}},
          Marshall C. Johnson\inst{\ref{osu}},
          William Cochran\inst{\ref{UTexas}},
          Hannah L. M. Osborne\inst{\ref{mullard},\ref{esogermany}},
          Rafael Brahm\inst{\ref{UniIbanexCile},\ref{MillenniumCile}},
          David~R.~Ciardi\inst{\ref{CaltechIPAC}},
          Karen A.\ Collins\inst{\ref{CFAharvard}},
          Mark~E.~Everett\inst{\ref{NOIRLab}},
          Steven Giacalone\inst{\ref{CaltechDep}},
          Eike W. Guenther\inst{\ref{Taut}},
          Artie Hatzes\inst{\ref{Taut}},
          Coel Hellier\inst{\ref{Keele}},
          Jonathan Horner\inst{\ref{UniSQ}}
          Petr Kab\'ath\inst{\ref{Ondrejov}},
          Judith Korth\inst{\ref{Lund}},
          Phillip MacQueen\inst{\ref{UTexas}},
          Thomas Masseron\inst{\ref{iac},\ref{ull}},
          Felipe Murgas\inst{\ref{iac},\ref{ull}},
          Grzegorz Nowak\inst{\ref{NicCopUni},\ref{iac},\ref{ull}},
          Joseph E. Rodriguez\inst{\ref{MSU}},
          Cristilyn N.\ Watkins\inst{\ref{CFAharvard}},
          Rob Wittenmyer\inst{\ref{UniSQ}},
          George Zhou\inst{\ref{UniSQ}},
          Carl Ziegler\inst{\ref{sfasu}},
          Allyson Bieryla\inst{\ref{CFAharvard}},
          Patricia T. Boyd\inst{\ref{goddard}},
          Catherine A. Clark\inst{\ref{JPLab},\ref{CaltechIPAC}},
          Courtney D. Dressing\inst{\ref{berkeley}},
          Jason D.\ Eastman\inst{\ref{CFAharvard}}
          Jan Eberhardt\inst{\ref{MaxPlanck}},
          Michael Endl\inst{\ref{UTexas}},
          Nestor Espinoza\inst{\ref{STScI}},
          Michael~Fausnaugh\inst{\ref{UTT}, \ref{MITKavli}},
          Natalia~M.~Guerrero\inst{\ref{MITKavli},\ref{UniFlorida}},
          Thomas Henning\inst{\ref{MaxPlanck}},
          Katharine Hesse\inst{\ref{MITKavli}},
          Melissa J.\ Hobson\inst{\ref{ObsGeneve}},
          Steve~B.~Howell\inst{\ref{AMES}},
          Andr\'es Jord\'an\inst{\ref{UniIbanexCile},\ref{MillenniumCile}},
          David W. Latham \inst{\ref{CFAharvard}},
          Michael B. Lund\inst{\ref{CaltechIPAC}},
          Ismael Mireles\inst{\ref{UNewMexico}},
          Norio Narita\inst{\ref{Komaba},\ref{AstroCenterTokyo},\ref{iac}},
          Marcelo Tala Pinto\inst{\ref{UniIbanexCile},\ref{MillenniumCile}},
          Teznie Pugh\inst{\ref{UTexas}},
          Samuel~N.~Quinn\inst{\ref{CFAharvard}},
          George Ricker\inst{\ref{MITKavli}},
          David~R.~Rodriguez\inst{\ref{STScI}}
          Felipe I.\ Rojas\inst{\ref{Pontificia},\ref{MillenniumCile}},
          Mark E. Rose\inst{\ref{AMES}},
          Alexander Rudat\inst{\ref{MITKavli}},
          Paula Sarkis\inst{\ref{MaxPlanck}},
          Arjun B. Savel\inst{\ref{umd}},
          Martin Schlecker\inst{\ref{arizona}},         
          Richard P. Schwarz\inst{\ref{CFAharvard}},
          Sara Seager \inst{\ref{MITKavli},\ref{MITEarth},\ref{MITaero}},
          Avi~Shporer\inst{\ref{MITKavli}},
          Jeffrey C. Smith\inst{\ref{SETI}},
          Keivan G. Stassun\inst{\ref{vanderbilt}},
          Chris Stockdale\inst{\ref{hazelwood}},
          Trifon Trifonov\inst{\ref{MaxPlanck},\ref{sofiauni},\ref{heidelberg}},
          Roland Vanderspek\inst{\ref{MITKavli}},
          Joshua N. Winn\inst{\ref{princeton}}
           \and
          Duncan Wright\inst{\ref{UniSQ}}          
}

\institute{INAF -- Osservatorio Astrofisico di Torino, Via Osservatorio 20, I-10025, Pino Torinese, Italy \label{INAF-oato}\\ 
              \email{ilaria.carleo@inaf.it}
        \and Instituto de Astrof\'{i}sica de Canarias (IAC), 38205 La Laguna, Tenerife, Spain \label{iac} 
         \and Departamento de Astrof\'isica, Universidad de La Laguna (ULL), E-38206 La Laguna, Tenerife, Spain \label{ull} 
         \and Sub-department of Astrophysics, Department of Physics, University of Oxford, Oxford, OX1 3RH, UK \label{phy_uni_oxford}
         \and Chalmers University of Technology, Department of Space, Earth and Environment, Onsala Space Observatory, SE-439 92 Onsala, Sweden. \label{OSO}
         \and Leiden Observatory, University of Leiden, PO Box 9513, 2300 RA, Leiden, The Netherlands\label{Leiden}
        \and Institute of Planetary Research, German Aerospace Center (DLR), Rutherfordstrasse 2, 12489 Berlin, Germany\label{DLR}
         \and INAF - Osservatorio Astrofisico di Catania, Via S. Sofia 78, I-95123, Catania, Italy\label{INAF-Catania}
         \and Dipartimento di Fisica, Universit\`a degli Studi di Torino, via Pietro Giuria 1, I-10125, Torino, Italy\label{UniTo}
         \and Department of Astronomy, The Ohio State University, 4055 McPherson Laboratory, 140 West 18th Avenue, Columbus, OH 43210 USA\label{osu}
         \and Mullard Space Science Laboratory, University College London, Holmbury St Mary, Dorking, Surrey RH5 6NT, UK\label{mullard}
         \and European Southern Observatory, Karl-Schwarzschild-Straße 2, D-85748 Garching bei Munchen, Germany\label{esogermany}
         \and Facultad de Ingeniera y Ciencias, Universidad Adolfo Ib\'{a}\~{n}ez, Av. Diagonal las Torres 2640, Pe\~{n}alol\'{e}n, Santiago, Chile\label{UniIbanexCile}
         \and Millennium Institute for Astrophysics, Chile\label{MillenniumCile}
         \and NASA Exoplanet Science Institute-Caltech/IPAC, Pasadena, CA 91125, USA\label{CaltechIPAC}
         \and Center for Astrophysics \textbar \ Harvard \& Smithsonian, 60 Garden Street, Cambridge, MA 02138, USA\label{CFAharvard}
         \and Department of Astronomy and McDonald Observatory, University of Texas at Austin, 2515 Speedway,~Stop~C1400,~Austin,~TX~78712,~USA\label{UTexas}
         \and NSF National Optical-Infrared Astronomy Research Laboratory, 950 N. Cherry Ave., Tucson, AZ 85719, USA\label{NOIRLab}
         \and Department of Astronomy, California Institute of Technology, Pasadena, CA 91125, USA\label{CaltechDep}
         \and Th\"uringer Landessternwarte Tautenburg, Sternwarte 5, D-07778 Tautenberg, Germany\label{Taut}
         \and Astrophysics Group, Keele University, Staffordshire, ST5 5BG, UK\label{Keele}
         \and Astronomical Institute, Czech Academy of Sciences, Fri\v{c}ova 298, 25165, Ond\v{r}ejov, Czech Republic\label{Ondrejov}
         \and Lund Observatory, Division of Astrophysics, Department of Physics, Lund University, Box 43, 22100 Lund, Sweden\label{Lund}
         \and Institute of Astronomy, Faculty of Physics, Astronomy and Informatics, Nicolaus Copernicus University, Grudzi\c{a}dzka 5, 87-100 Toru\'n, Poland\label{NicCopUni}
         \and Center for Data Intensive and Time Domain Astronomy, Department of Physics and Astronomy, Michigan State University, East Lansing, MI 48824, USA\label{MSU}
         \and University of Southern Queensland, Centre for Astrophysics, UniSQ Toowoomba, West Street, QLD 4350, Australia\label{UniSQ}
         \and Department of Physics, Engineering and Astronomy, Stephen F. Austin State University, 1936 North St, Nacogdoches, TX 75962, USA\label{sfasu}
         \and NASA Goddard Space Flight Center, Greenbelt, MD 20771, USA\label{goddard}
         \and Jet Propulsion Laboratory, California Institute of Technology, Pasadena, CA 91109 USA\label{JPLab}
         \and Department of Astronomy, University of California Berkeley, Berkeley, CA 94720, USA\label{berkeley}
         \and Max-Planck-Institut für Astronomie, Königstuhl 17, D-69117 Heidelberg, Germany\label{MaxPlanck}
         \and Space Telescope Science Institute, 3700 San Martin Drive, Baltimore, MD 21218, USA\label{STScI}
         \and Department of Physics \& Astronomy, Texas Tech University, Lubbock TX, 79410-1051, USA\label{UTT} 
         \and Department of Physics and Kavli Institute for Astrophysics and Space Research, Massachusetts Institute of Technology, 77 Massachusetts Avenue, Cambridge, MA 02139, USA\label{MITKavli}
         \and Department of Astronomy, University of Florida, Gainesville, FL 32611, USA\label{UniFlorida}
         \and Observatoire de Gen\`eve, D\'epartement d'Astronomie, Universit\'e de Gen\`eve, Chemin Pegasi 51b, 1290 Versoix, Switzerland\label{ObsGeneve}
         \and NASA Ames Research Center, Moffett Field, CA 94035, USA\label{AMES}
         \and Department of Physics and Astronomy, University of New Mexico, 210 Yale Blvd NE, Albuquerque, NM, USA\label{UNewMexico}
         \and Komaba Institute for Science, The University of Tokyo, 3-8-1 Komaba, Meguro, Tokyo 153-8902, Japan\label{Komaba}
         \and Astrobiology Center, 2-21-1 Osawa, Mitaka, Tokyo 181-8588, Japan \label{AstroCenterTokyo}
         \and Instituto de Astrof\'isica, Facultad de F\'isica, Pontificia Universidad Cat\'olica de Chile, Chile\label{Pontificia}
         \and Department of Astronomy, University of Maryland, College Park, College Park, MD 20742 USA\label{umd}
         \and Department of Astronomy/Steward Observatory, The University of Arizona, 933 North Cherry Avenue, Tucson, AZ 85721, USA\label{arizona}
         \and Department of Earth, Atmospheric and Planetary Sciences, Massachusetts Institute of Technology, 77 Massachusetts Avenue, Cambridge, MA 02139, USA\label{MITEarth} 
         \and Department of Aeronautics and Astronautics, Massachusetts Institute of Technology, 77 Massachusetts Avenue, Cambridge, MA 02139, USA\label{MITaero}
         \and SETI Institute, 339 N Bernardo Ave Suite 200, Mountain View, CA 94043, USA\label{SETI}
         \and Department of Physics \& Astronomy, Vanderbilt University, Nashville, TN, USA\label{vanderbilt}
         \and Hazelwood Observatory, Australia\label{hazelwood}
         \and Department of Astronomy, Sofia University ``St Kliment Ohridski'', 5 James Bourchier Blvd, BG-1164 Sofia, Bulgaria\label{sofiauni}
         \and Landessternwarte, Zentrum f\"ur Astronomie der Universit\"at Heidelberg, K\"onigstuhl 12, D-69117 Heidelberg, Germany\label{heidelberg}
         \and CAS Key Laboratory of Planetary Sciences, Purple Mountain Observatory, Chinese Academy of Sciences, Nanjing 210008, China\label{CASKLab}
         \and Department of Astrophysical Sciences, Princeton University, Princeton, NJ 08544, USA\label{princeton}
         \\
}

\date{Received Date Month YYYY; accepted Date Month YYYY}

 
  \abstract
   {Hot and warm Jupiters might have undergone the same formation and evolution path, but the two populations exhibit different distributions of orbital parameters, challenging our understanding on their actual origin. }
   {The present work, which is the results of our warm Jupiters survey  carried out with the CHIRON spectrograph within the KESPRINT collaboration, aims to address this challenge by studying two planets that could help bridge the gap between the two populations.}
   {We report the confirmation and mass determination of a hot Jupiter (orbital period shorter than 10 days), TOI-2420\,b, and a warm Jupiter, TOI-2485\,b. We performed a joint analysis using a wide variety of spectral and photometric data in order to characterize these planetary systems.}
   {We found that TOI-2420\,b has an orbital period of P$_{\rm b}$=5.8 days, a mass of M$_{\rm b}$=0.9 M$_{\rm J}$ and a radius of R$_{\rm b}$=1.3 R$_{\rm J}$, with a planetary density of 0.477 \gc; while TOI-2485\,b has an orbital period of P$_{\rm b}$=11.2 days, a mass of M$_{\rm b}$=2.4 M$_{\rm J}$ and a radius of R$_{\rm b}$=1.1 R$_{\rm J}$ with density 2.36 \gc.   }
   {With current parameters, the migration history for TOI-2420\,b and TOI-2485\,b is unclear: the high-eccentricity migration scenarios cannot be ruled out, and TOI-2485\,b's characteristics may rather support this scenario.}

   \keywords{}

\titlerunning{Mass determination of two Jupiter-sized planets orbiting slightly evolved stars: TOI-2420\,b and TOI-2485\,b}
\authorrunning{Carleo et al.}
   \maketitle
%

 \section{Introduction} \label{sec:intro}
Almost 30 years since the discovery of the first hot Jupiter (HJ), 51~Peg~b \citep{Mayor1995}, the formation and migration history of close-in giant planets is still debated. The orbit of 51 Peg b proved a huge surprise to astronomers at the time \citep[see e.g.][who state that the newly discovered planet `is surely the most problematic find in recent memory']{Guillot96}. A planet comparable in mass to Jupiter moving on an orbit extremely close to its host star ran counter to the prevailing understanding of planet formation at the time, which was based solely on our knowledge of the Solar system\footnote{For a detailed overview of our knowledge of the Solar system, and a discussion of how it has influenced our understanding and knowledge of planet formation, we direct the interested reader to \cite{SSRev}, and references therein; the review by \cite{Lissauer93} describes our understanding of planet formation in the years before the dawn of the Exoplanet Era.}. 

51 Peg b was the first of a population of planets that quickly became known as the `hot Jupiters' (HJs) \citep[e.g.][]{Sch96} - giant planets orbiting their host stars with periods less than around ten days . In the years that followed, and more exoplanets were discovered, a number of `warm Jupiters' (WJs) were also found - giant planets with orbital periods between 10 and 200 days \citep[e.g.,][]{Dawson2018} - again, dramatically different to the planets in our own Solar system.
The origin of both HJs and WJs has been heavily debated. 

Both populations could have originated through in-situ formation \citep{1997Sci...276.1836B}, disk migration \citep{Lin1986} or high eccentricity migration \citep{Wu2003}. However, the two populations present differences in some of their properties:

\noindent\textit{a)} the occurrence rate of WJs per log interval of period is lower than that for HJs (see histogram in Fig. 4 of \citealt{Dawson2018}), but the total occurrence rate of WJs is larger (i.e., \citealt{Wittenmyer2010, Zink2023}); 

\noindent\textit{b)} most HJs present low eccentricities, while WJs present a wide range of eccentricities (i.e., \citealt{Correia2020,Zink2023}); 

\noindent\textit{c)} HJs generally lack nearby companions, while WJs have been found with nearby super-Earths \citep{Huang2016}, even though recent studies have demonstrated that a fraction of HJs $\geqslant$12$\pm$6\% have nearby small (1-4 R$_{\oplus}$) companions \citep{Wu2023} and $\sim$30\% of HJs have at least one Warm/Cold Jupiter companion \citep{Zink2023}. 

These differences are likely related to the formation site and migration history of the planets involved. For example, disk migration is thought to be the primary mechanism that produces HJs, but cannot explain the wide eccentricity distribution of WJs. On the other hand, WJs might have experienced high-eccentricity tidal migration, but this mechanism is more efficient for closer WJs, since the tidal dissipation has a strong dependence with the semi-major axis. It is thus important to study  hot and warm Jupiters and assess the relative effectiveness of the different formation scenarios proposed for these planets. For a more comprehensive overview of the different theories on formation and evolution of close-in giant planets, as well as of the similarities and dissimilarities of hot and warm Jupiters see Sec. 4.3 in \citealt{Dawson2018}.


In the past few years, the Transiting Exoplanet Survey Satellite (TESS; \citealt{Rickeretal2014}) released thousands of planetary candidates and the exoplanetary community have put substantial effort into the radial velocity (RV) follow-up with ground based spectrographs in order to confirm the planetary nature and determine the mass of the candidates. With this effort, many hot and warm Jupiters have been confirmed (144 in total), allowing us to greatly improve the statistical significance of our sample, and thus improve our understanding of the difference between these two populations.

In this paper, we present the mass determination of two close-in giant planets, one HJ, TOI-2420\,b, and one WJ, TOI-2485\,b. We present the observations of the two targets, including TESS photometry, ground-based photometry, and spectroscopy in Section \ref{sec:obs}, the stellar characterization in Section \ref{Subsection: Stellar modelling}, the planetary systems' modelling with the transit and RV joint fit in  analysis of the photometry together with the transit fit and RV modeling in Section \ref{sec:planet}. Finally, we discuss our results and present our conclusions in Section \ref{sec:disc}.

 \section{Observations} \label{sec:obs}

\subsection{TESS photometry} 
TOI-2420 (TIC 268532343) was observed by TESS  between 2018 September 20 and 2019 January 24 in sector 3 on CCD1 of Camera 1, as well as between 2020 September 23 and 2020 November 20 in sector 30 on CCD1 of Camera 1, and was alerted on 2020 November 25. TOI-2485  (TIC 328934463) was observed between 2020 March 19 and 2020 May 04 in sector 23 on CCD4 of Camera 2, and between 2022 March 26 and 2022 May 11 in sector 50 on CCD3 of Camera 2, and alerted on 2021 February 11. The data taken in each sector were observed in 30 min, 10 min, 30 min and 2 min cadence, respectively. The data were reduced by both the MIT Quick-Look Pipeline (QLP; \citealt{Huangetal2020,Kunimotoetal2021}), and the TESS Science Processing Operations Center (SPOC; \citealt{Jenkinsetal2010}) pipeline. The SPOC pipeline was adapted from the Kepler mission pipeline at NASA Ames Research Center. The pipeline uses simple aperture photometry (SAP; \citealt{Twickenetal2010}) to produce time series light curves. A further presearch data conditioning (PDCSAP) algorithm was subsequently used to correct for common instrumental systematics in the data \citep{Stumpeetal2012,Smithetal2012}. For the sector 50 short cadence data of TOI-2485, we downloaded the SPOC light curve from the Mikulski Archive for Space Telescopes (MAST\footnote{\url{https://mast.stsci.edu/}}). For the other data, we downloaded light curves extracted from the TESS-SPOC pipeline \citep{Caldwelletal2020}, which followed the same reduction routines as SPOC but were processed from TESS full frame images. 

Transit searches and signal assessments were performed by both the SPOC and the QLP pipelines. The light curves were further analysed using the transit search algorithm, DST \citep[D\'etection Sp\'ecialis\'ee de Transits; ][]{2012A&A...548A..44C}. In QLP and DST pipelines, a transit signal was detected in the TOI-2420 data with period $P = 5.84115 \pm 0.00257$ days, epoch $T_{0,BTJD} = 1388.41352 \pm 0.00280$ (where BTJD is defined as BJD-7000), transit duration $T_{14} = 4.33 \pm 0.16$ h and transit depth $df = 0.3023 \pm 0.0185$ \%. In the TOI-2485 light curves, a transit signal with $P = 11.23702 \pm 0.00654$ days, $T_{0,BTJD} = 1939.78211 \pm 0.00327$, $T_{14} = 6.94 \pm 0.19$ h and $df = 0.5328 \pm 0.0231$ \% was detected. 

We iteratively searched for further transit signatures in both datasets after transit signals of the first planet candidates were filtered out. There were no additional transiting candidates detected in both systems. Also no clear periodic variability is found in the TESS light curves. 

\subsection{Ground-based Photometry\label{subsec:ground_phot}}

The TESS pixel scale is $\sim 21\arcsec$ pixel$^{-1}$ and photometric apertures typically extend out to roughly 1 arcminute, generally causing multiple stars to blend in the TESS photometric aperture. To attempt to determine the true source of the TESS detection, we acquired ground-based time-series follow-up photometry of the fields around TOI-2420 and TOI-2485 as part of the TESS Follow-up Observing Program \citep[TFOP;][]{collins:2019}\footnote{https://tess.mit.edu/followup}. We used the {\tt TESS Transit Finder}, which is a customized version of the {\tt Tapir} software package \citep{Jensen:2013}, to schedule our transit observations.

\subsubsection{WASP}
WASP-South (Wide-Angle Search for Planets) was the southern station of the WASP transit-search survey \citep{2006PASP..118.1407P}, and consisted of an array of 8 wide-field cameras observing fields with a typical 10-min cadence. The field of TOI-2420 was observed over spans of 160 to 180 nights in each year from 2006 to 2011. In all, 21\,150 photometric observations were obtained, using a 48" extraction aperture within which TOI-2420 is the only bright star. 

While TOI-2420b was not a WASP candidate with hindsight we notice that the standard WASP transit-search algorithm finds the 0.3\%-deep transit and reports an ephemeris of:
\[ {\rm TDB(JD)} = 245\,4432.934 \pm\ 0.012 + N \times 5.84265 \pm\ 0.00014.\] 

\begin{figure}
\includegraphics[width=9cm]{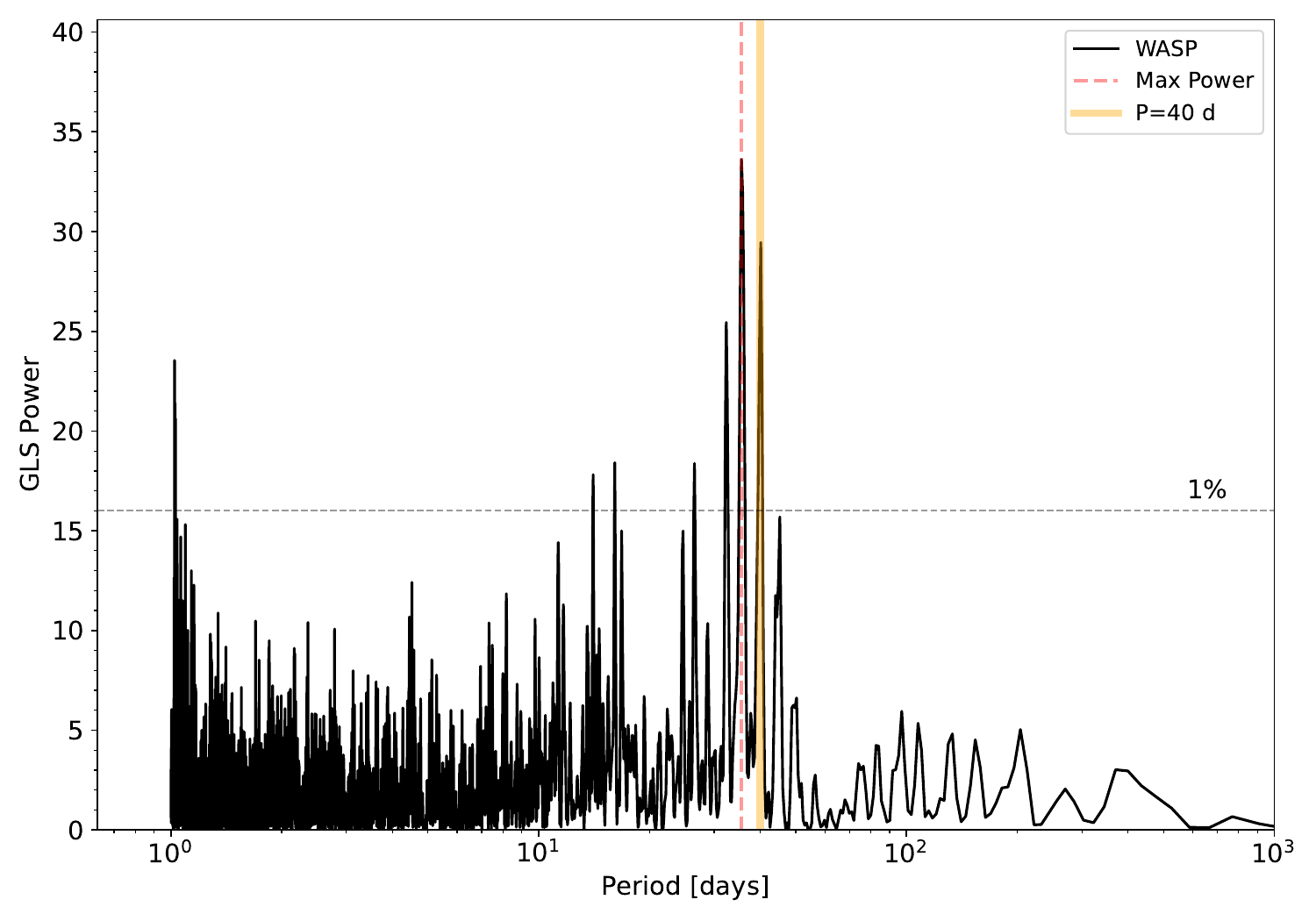}
  \caption{GLS periodogram of the WASP-South data for TOI-2420 from 2006 to 2011. There is a possible signal near 40 d, along with aliases from the yearly sampling. The dotted horizontal line  is the 1\%-likelihood false-alarm level.}
\label{fig:wasp}
\end{figure}

We also searched the WASP lightcurve for any rotational modulation.
We computed the generalised Lomb-Scargle (GLS) periodograms \citep{zech09} and estimated the false alarm probability (FAP) via a bootstrap method \citep{Murdoch1993,Hatzes2016} that generates 1,000 artificial photometric datasets obtained from the real data, making random permutations in the photometry values. We found the maximum period at $\sim$ 36 days with a FAP lower than 10$^{-6}$  (see Fig.~\ref{fig:wasp}). 

\subsubsection{LCOGT\label{subsubsec:lcogt}}

We observed a partial transit window of the planet candidate TOI-2420.01 in Sloan $i'$ on UTC 2020 December 11 from the Las Cumbres Observatory Global Telescope (LCOGT) \citep{Brown:2013} 1\,m network node at Cerro Tololo Inter-American Observatory in Chile (CTIO). We also observed a full transit window in alternating Sloan $g'$ and Sloan $i'$ on UTC 2021 September 29 from another LCOGT 1\,m network node at McDonald Observatory near Fort Davis, Texas, United States (McD). The 1\,m telescopes are equipped with a $4096\times4096$ SINISTRO camera having an image scale of $0\farcs389$ per pixel, resulting in a $26\arcmin\times26\arcmin$ field of view and the images were calibrated by the standard LCOGT {\tt BANZAI} pipeline \citep{McCully:2018}, and differential photometric data were extracted using {\tt AstroImageJ} \citep{Collins2017}. We used circular photometric apertures with radius $7\farcs0$. The target star aperture excluded all of the flux from the nearest known neighbor in the \textit{Gaia} DR3 catalog (\textit{Gaia} DR3 2356241534150962944), which is $\sim49\arcsec$ south of TOI-2420. The light curve data are available on the {\tt EXOFOP-TESS} website\footnote{\href{https://exofop.ipac.caltech.edu/tess/target.php?id=268532343}https://exofop.ipac.caltech.edu/tess/target.php?id=268532343} and are included in the global modelling described in section \ref{sec:planet}.

\subsubsection{KeplerCam}

We observed a partial transit window of the planetary candidate TOI-2485.01 in Sloan $i'$ on UTC 2021 April 17 from KeplerCam, which is installed on the 1.2\,m telescope at the \textit{Fred Lawrence Whipple} Observatory. The $4096\times4096$ Fairchild CCD 486 detector has an image scale of $0\farcs672$ per $2\times2$ binned pixel, resulting in a $23\farcm1\times23\farcm1$ field of view. The images were calibrated and photometric data were extracted with {\tt AstroImageJ} using a circular aperture with radius 6$\farcs$7. The target star aperture excluded all of the flux from the nearest known neighbor in the \textit{Gaia} DR3 catalog (\textit{Gaia} DR3 1443530261849361152), which is $\sim16\arcsec$ north of TOI-2485. The light curve data are available on the {\tt EXOFOP-TESS} website\footnote{\href{https://exofop.ipac.caltech.edu/tess/target.php?id=328934463}https://exofop.ipac.caltech.edu/tess/target.php?id=328934463} and are included in the global modelling described in section \ref{sec:planet}.

\subsection{Ground-based Spectroscopy\label{subsec:ground_spec}}
We collected RVs with different ground-based instruments. The RV data described in the following subsections are listed for both targets in Tables \ref{tab:TOI2420_rv} and \ref{tab:TOI2485_rv}.

\subsubsection{CHIRON}
We observed TOI-2420 and TOI-2485 with the spectrograph CHIRON at SMARTS 1.5-meter telescope at Cerro Tololo Inter-American Observatory, Chile \citep{2013PASP..125.1336T}, within the large observing program (ID: CARL-20B-3081, PI: Carleo) aimed to survey a sample of $\sim$20 warm Jupiters carried out within the KESPRINT collaboration\footnote{\url{www.kesprint.science}.} \citep[i.e.,][]{deLeon2021,Smith2022,Tran2022,Kabath2022,Korth2023}, which aims to confirm and characterize planet candidates from the \textit{K2} and TESS space missions. The CHIRON observations were performed in Slicer mode, reaching a spectral resolving power of $R=80,000$ over the wavelength range of 4100 to $8700$\,\AA{}. 
We collected 18 CHIRON spectra for TOI-2420 and 14 spectra for TOI-2485. The data reduction was performed through the official spectral extraction pipeline of CHIRON \citet{2021AJ....162..176P}. Radial velocities were obtained via a least-squares deconvolution of the observation against a synthetic non-rotating ATLAS9 model atmosphere spectrum \citep{Castelli:2004}. The least-squares deconvolution kernel was modeled via a broadening kernel in order to include the effects of radial velocity shift, rotational, instrumental, and macroturbulent broadening \citep{2021AJ....161....2Z}. The average RV precision obtained for TOI-2420 is 21\,m~s$^{-1}$ and 19\,m~s$^{-1}$ for TOI-2485. 

\begin{figure}
  \centering
  \includegraphics[width=0.50\textwidth, trim=0 8cm 0 0]{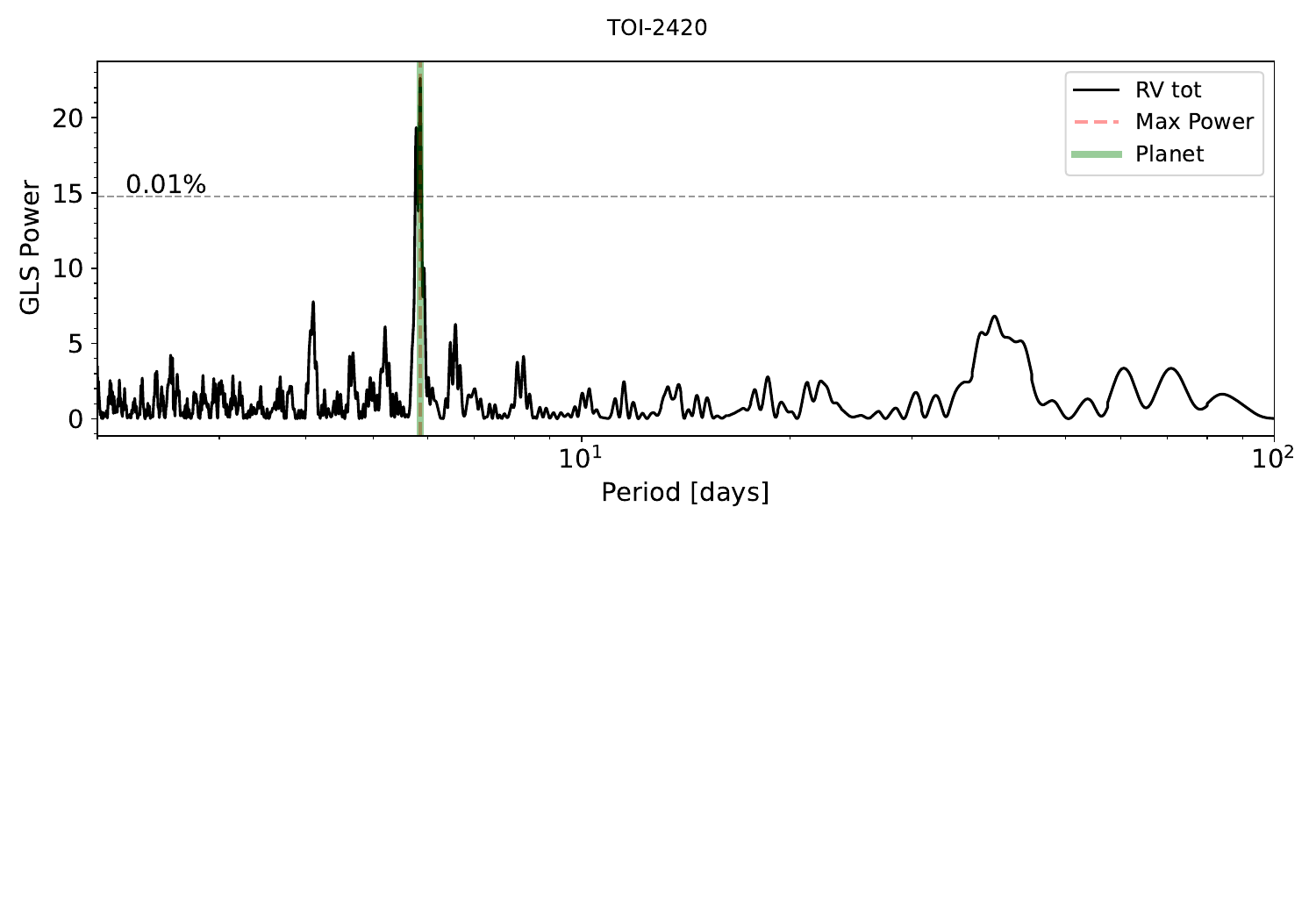}
    \includegraphics[width=0.50\textwidth, trim=0 8cm 0 0]{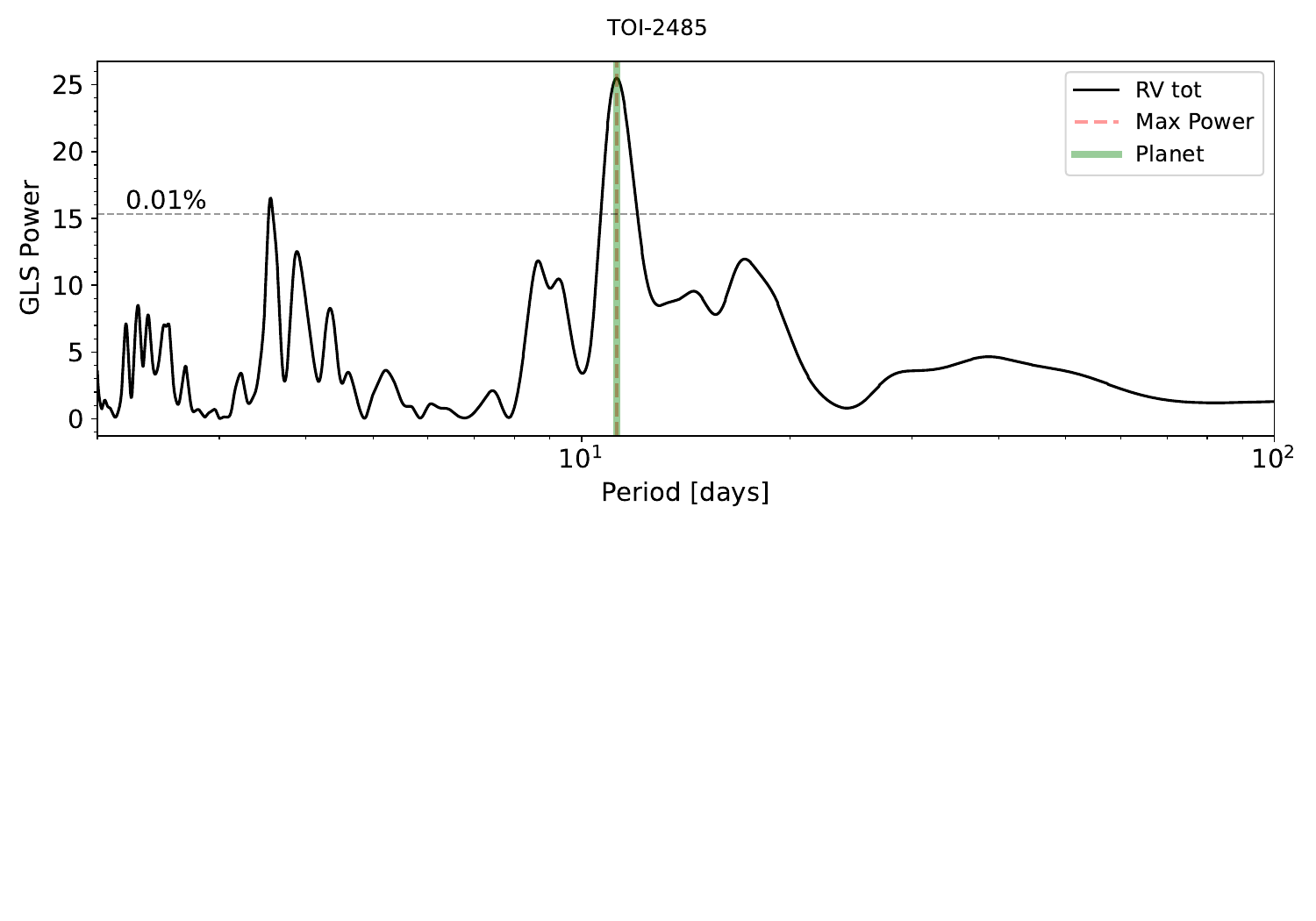}
  \caption{Periodograms of the RVs data for TOI-2420 (upper panel) and TOI-2485 (lower panel). The dotted horizontal line represent the 0.01\% false-alarm level, while the green vertical line is the maximum power, which corresponds to the planetary period. \label{fig:periodograms}}
\end{figure}

We computed the GLS periodograms for both targets (Fig. \ref{fig:periodograms}). They exhibit a highly significant periodicity at 5.8 days and 11.2 days, for TOI-2420 and TOI-2485, respectively, which correspond to the planetary signals. The resulting FAP estimated via bootstrap is lower than 10$^{-6}$.

\subsubsection{Minerva-Australis} 
TOI-2420 was observed between 2021 June 5 and 2021 October 4 using the 
MINERVA-Australis telescope array \citep{addison19}, located at Mt. 
Kent Observatory, Australia. Minerva-Australis is an array of four 
identical 0.7 m telescopes linked via fiber feeds to a single KiwiSpec 
echelle spectrograph at a spectral resolving power of $R\sim$80,000 over 
the wavelength region of 5000-6300\AA. The array is wholly dedicated to 
radial-velocity follow-up of TESS planet candidates (e.g., \citealt{nielsen19, addison21, wittenmyer22, cargoship, clark23}). Two simultaneous 
fibres provide wavelength calibration and correct for instrumental 
variations.  The calibration fibres are illuminated by a quartz lamp 
through an iodine cell, eliminating contamination by saturated Argon lines.  
Radial velocities for the observations are derived for each telescope from 
a least-squares deconvolution against a synthetic non rotating template, 
similar to the CHIRON pipeline.  Each epoch consists of 30-60 minute 
exposures from up to four individual telescopes.  Fibres 3, 4, 5, and 6 
obtained 45, 16, 6, and 37 epochs respectively.  The radial velocities 
from each telescope are treated as coming from separate instruments here 
to account for small velocity offsets between the fibres.

\subsubsection{Tull Coud\'{e} Spectrometer}
We observed TOI-2420 with the Tull Coud\'{e} Spectrometer (TS23) \citep{1995PASP..107..251T} of the McDonald Observatory 2.7m \textit{Harlan J. Smith} Telescope.   TS23 is a cross-dispersed echelle white-pupil spectrograph with a Tektoronix $2048\times2048$ CCD detector.  A 1.2 arcsec wide slit gave spectral resolving power $R={\lambda}/{\delta \lambda} = 60,000$.  We focus the stellar image onto the slit with a wave-front sensor. This instrumental configuration gives complete spectral coverage from 3400\,{\AA} to 5800\,{\AA}, and then increasingly large inter-order gaps exist out to 10,800\,{\AA}.  We insert an I$_2$ gas absorption cell in front of the spectrograph entrance slit in order to impose a stable set of fixed absorption lines on the stellar spectrum before it enters the spectrograph.  This enables us to measure precise radial velocity variations of the target star with respect to the I$_2$ lines \citep[cf.][]{1996PASP..108..500B}.  At the start of each night, the spectrograph is automatically re-positioned to within 0.2 pixels of a standard reference position.  We obtained 21 separate visits to TOI-2420 between 2021 July 18 and 2022 December 13.   We used an exposure meter to terminate each exposure level at a preset signal to noise level.  The exposure meter data are then used to compute an accurate flux-weighted barycentric velocity correction for each spectrum. All of the CCD frames were reduced and the echelle spectra were extracted using a script of standard IRAF procedures.  We then computed the radial velocities from the extracted spectra using the AUSTRAL code \citep{2000A&A...362..585E}.

\subsubsection{FEROS} 
TOI-2485 was monitored with the The Fiber-fed Extended Range Optical Spectrograph \citep[FEROS]{feros} mounted to the MPG2.2m telescope at the ESO La Silla Observatory, in Chile. FEROS has a spectral resolution of $R=48,000$ and uses a second fibre to trace instrument induced spectral displacements. The observations of TOI-2485 were obtained in the context of the Warm gIaNts with tEss (WINE) collaboration which focuses on the systematic discovery of transiting warm Jupiters \citep{brahm:2019,jordan:2020,brahm:2020,schlecker:2020,hobson:2021,trifonov:2021,trifonov:2023,bozhilov:2023,brahm:2023,hobson:2023,eberhardt:2023,jones:2024}. We obtained 15 FEROS spectra between February of 2021 and July of 2023 using an exposure time of 1200 s obtaining spectra with signal-to-noise ratios between 70 and 110 per resolution element depending on the weather and observing conditions. FEROS data were proccessed with the \texttt{ceres} pipeline which generates as final outputs the two dimensional spectrum, and the determination of precise radial velocities and bisector span measurements using the cross-correlation technique. The mean error of these radial velocity measurements was of 9 m/s. \texttt{ceres} performs also a rough estimation of the stellar parameters, and for the case of TOI-2485, we obtained \teff=5900 $\pm$ 100 K, \logg=4.2 $\pm$ 0.2 dex \feh = 0 $\pm$ 0.1, and \vsini = 5 $\pm$ 1 \kms.

\subsubsection{TRES}
TOI-2485 was observed 11 times from UT 2021 February 17 to February 27 using the Tillinghast Reflector Echelle Spectrograph \citep[TRES;][]{furesz:2008}\footnote{\url{http://www.sao.arizona.edu/html/FLWO/60/TRES/GABORthesis.pdf}} on the 1.5m Tillinghast Reflector at the \textit{Fred L. Whipple} Observatory (FLWO) on Mt. Hopkins, AZ. TRES has a resolution of R = 44,000, and covers a spectral wavelength range of 3850-9096\AA. The reduction process for TRES is described in detail in  \citet{Buchhave:2010}, and the RV extraction process using a median combined template is presented in \citet{Quinn:2012}. To better understand the host star parameters, the spectra were analyzed using the Stellar Parameter Classification (SPC) package \citep{Buchhave2012} providing a comparison constraint on the \feh, \teff, and rotational velocity of TOI-2485 of 0.005$\pm$0.008 dex, 5982$\pm$50 K, and 6.01$\pm$0.05 \kms.

\subsection{High Resolution Imaging}
    As part of our standard process for validating transiting exoplanets to assess the the possible contamination of bound or unbound companions on the derived planetary radii \citep{ciardi2015}, we observed TOI~2420 and TOI~2485 with optical speckle observations at SOAR and WIYN and near-infrared adaptive optics (AO) imaging at Palomar and Lick Observatories.
	
	\subsubsection{Optical Speckle Imaging}

        We searched for stellar companions to TOI-2420 and TOI-2485 with speckle imaging on the 4.1-m Southern Astrophysical Research (SOAR) telescope \citep{tokovinin2018} on UT 2020 December 3  and 2021 February 27, respectively, observing in Cousins I-band, a similar visible bandpass as TESS. This observations were both sensitive to a 5.0-magnitude fainter star at an angular distance of 1 arcsec from the target. More details of the observations within the SOAR TESS survey are available in \citet{ziegler2020}. No nearby stars were detected within 3\arcsec of either TOI-2420 or TOI-2485 in the SOAR observations.
 
	\subsubsection{NESSI} 

\begin{figure}
    \centering
    \includegraphics[width=0.4\textwidth]{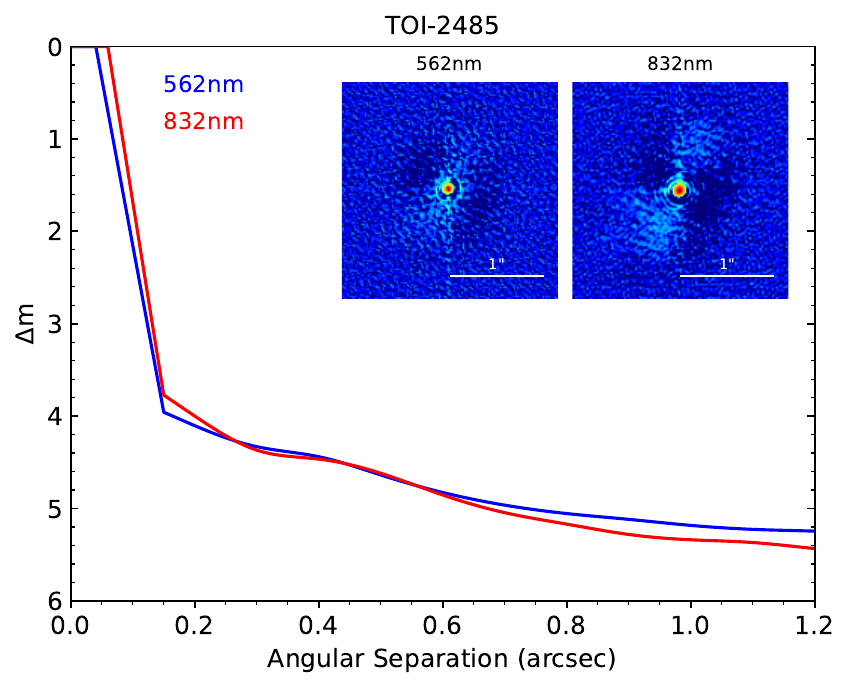}
    \caption{NESSI speckle imaging results from observations of TOI-2485 on 2021 April 1. Sensitivity curves and reconstructed images are shown for each filter (central wavelengths 562 and 832~nm). No nearby companions have been detected.
    }
    \label{fig:nessi}
\end{figure}

 We observed TOI-2485 on UT 2021 April 1 and UT 2022 April 18 using the NN-EXPLORE Exoplanet Stellar Speckle Imager (NESSI; \citealt{Scott2018}), a speckle imager employed at the WIYN 3.5~m telescope on Kitt Peak.  NESSI was used to obtain simultaneous speckle imaging in two filters with central wavelengths $\lambda_c = 562$ and 832~nm for the 2021 observation, but only the 832~nm was available for the 2022 observation. Each observation consisted of a set of 9 1000-frame 40~ms exposures.  NESSI's field-of-view was limited to a $256\times256$ pixel sub-array readout, resulting in a $4.6\times4.6$~arcsecond field.  However, our speckle measurements were further confined to an outer radius of 1.2~arcseconds from the target star.  Speckle imaging of a point source standard star was taken in conjunction to each observation of the TOI.   The standard observation consisted of a single 1000-frame image set and was used to calibrate the intrinsic PSF.
These speckle data were reduced using the pipeline process described in \cite{howell2011}.  Among the pipeline products are reconstructed images of the field around TOI-2485 in each filter.  We used these to measure contrast curves, setting detection limits on point sources close to the TOI.  No companion sources were detected for TOI-2485 (Fig.~\ref{fig:nessi}).\\

	\subsubsection{Near-Infrared AO Imaging}
    
\begin{figure}
    \centering
    \includegraphics[width=0.4\textwidth]{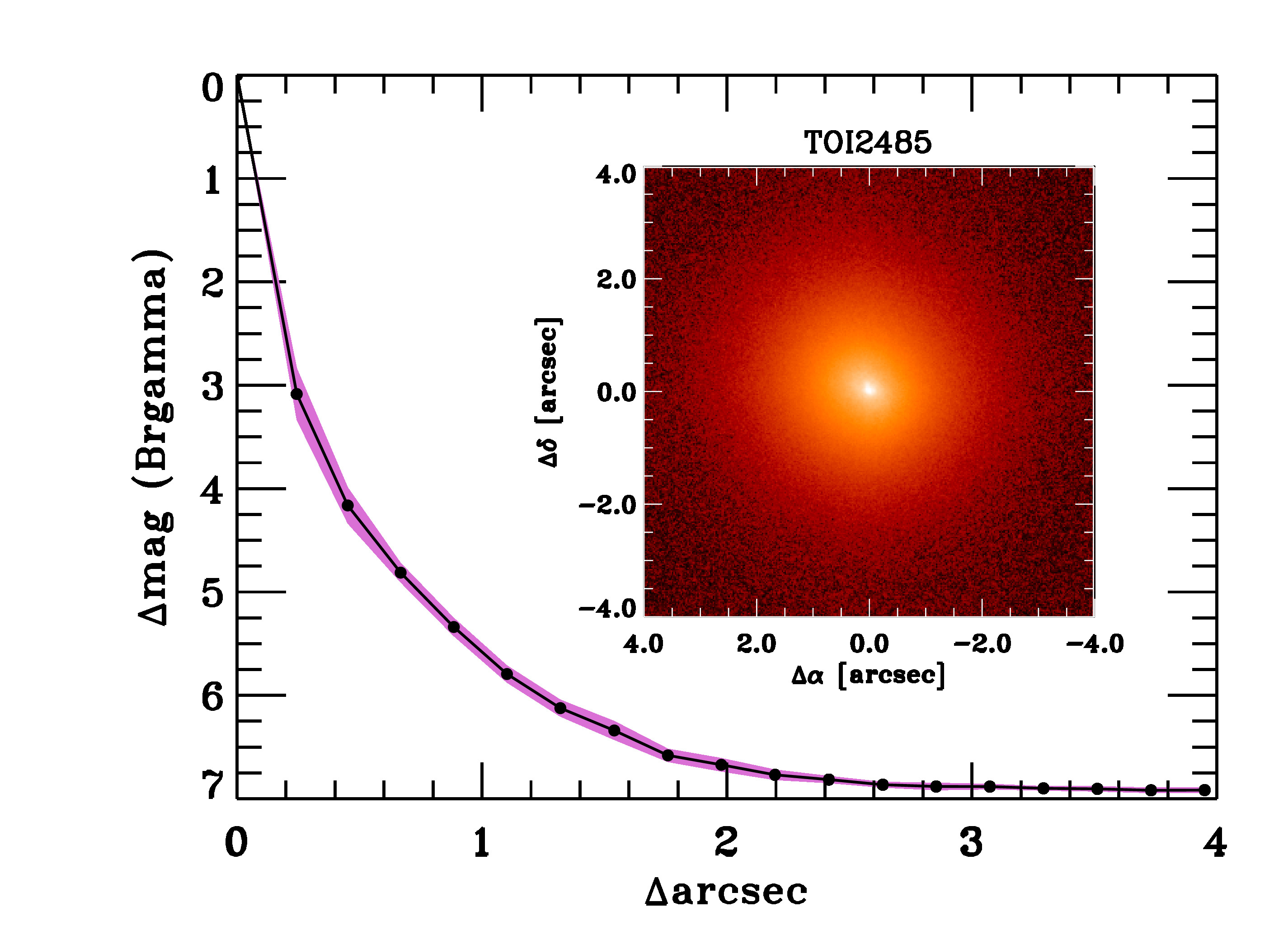}
    \caption{NIR AO imaging and sensitivity curves for the Paloamr Observations of TOI-2485. {\it Inset:} Image of the central portion of the image. No nearby companions have been detected.
    }
    \label{fig:palomar_aoimaging}
\end{figure}

	Observations of TOI-2485 were made on UT 2023 June 7 with the PHARO instrument \citep{hayward2001} on the Palomar Hale (5m) behind the P3K natural guide star AO system \citep{dekany2013} in the narrowband Br-$\gamma$ filter $(\lambda_o = 2.1686; \Delta\lambda = 0.0326~\mu$m). The PHARO pixel scale is $0.025\arcsec$ per pixel. A standard 5-point quincunx dither pattern with steps of 5\arcsec\ was repeated twice with each repeat separated by 0.5\arcsec. The reduced science frames were combined into a single mosaiced image with a final resolutions of 0.21\arcsec.  The sensitivity of the final combined AO image were determined by injecting simulated sources azimuthally around the primary target every $20^\circ $ at separations of integer multiples of the central source's FWHM \citep{furlan2017}. The brightness of each injected source was scaled until standard aperture photometry detected it with $5\sigma $ significance.  The final $5\sigma $ limit at each separation was determined from the average of all of the determined limits at that separation and the uncertainty on the limit was set by the rms dispersion of the azimuthal slices at a given radial distance.  The Palomar sensitivities are shown in (Fig.~\ref{fig:palomar_aoimaging}).  

	\subsubsection{ShARCS} 
    TOI-2485 was observed on UT 2021 March 04 using the ShARCS camera on the Shane 3-meter telescope at Lick Observatory \citep{2012SPIE.8447E..3GK, 2014SPIE.9148E..05G, 2014SPIE.9148E..3AM}. Observations were taken with the Shane adaptive optics system in natural guide star mode in order to search for nearby, unresolved stellar companions. Sequences of observations were collected using a $K_s$ filter ($\lambda_0 = 2.150$ $\mu$m, $\Delta \lambda = 0.320$ $\mu$m) and a $J$ filter ($\lambda_0 = 1.238$ $\mu$m, $\Delta \lambda = 0.271$ $\mu$m). The data were reduced using the publicly available \texttt{SImMER} pipeline \citep{2020AJ....160..287S, 2022PASP..134l4501S}.\footnote{https://github.com/arjunsavel/SImMER} No stellar companions were found within detection limits. We refer the reader to Dressing et al. (in prep) for more information about these observations.

\section{Stellar modelling} \label{Subsection: Stellar modelling}
\subsection{Spectroscopic modelling of the host stars}
We carried out spectroscopic modelling of the two exoplanet host stars using our co-added CHIRON spectra with the Spectroscopy Made Easy\footnote{\url{http://www.stsci.edu/~valenti/sme.html}} 
 \citep[SME;][]{vp96, pv2017} version 5.2.2.
This software fits spectral observations to synthetic spectra computed with  atomic and molecular line data     from VALD\footnote{\url{http://vald.astro.uu.se}} \citep{Ryabchikova2015} and different stellar atmosphere grids for a chosen set of parameters. We  used the  Atlas12     
  \citep{Kurucz2013} atmospheric model for both host stars.   
 A more detailed description of the SME  modelling can be found in 
  \citep{2018A&A...618A..33P}. In summary,  
we fitted spectral lines   sensitive to  different parameters: the line wings of H$\alpha$ at 6\,563\AA~to model \teff, and the line wings of the \ion{Ca}{i} lines at 6\,102~\AA, 6\,122~\AA, and 6162~\AA~for \logg. 
The abundances of iron, calcium, and sodium, and the projected rotational velocity ($V \sin i_\star$), were fitted to narrow and unblended spectral lines between 5\,900~\AA~and 6\,600~\AA. As a final check of the model, we fitted the Na doublet at 5888~\AA~and 5895~\AA, sensitive to both gravity and effective temperature. 
We fixed the 
micro-turbulent velocity, $V_{\rm mic}$ 
to 1~km~s$^{-1}$ \citep{bruntt08} for both host stars,   
and the macro-turbulent velocity, $V_{\rm mac}$  to  4.1~km~s$^{-1}$  for TOI-2420 and 4.4~km~s$^{-1}$ for TOI-2485 \citep{Doyle2014}.

All SME results for both host stars are listed in Table~\ref{Table: stellar spectroscopic parameters} which are adopted as the  final spectroscopic parameters.

The surface gravities combined with the effective temperatures suggests a G7\,IV and G0\,IV spectral type for TOI-2420 and TOI-2485, respectively \citep{2013ApJS..208....9P}. 
 
\subsection{Modelling of stellar masses and radii} \label{Subsection: stellar mass and radius}
\begin{figure*}
     \centering
     \begin{subfigure}[b]{0.49\textwidth}
         \centering
         \includegraphics[width=\textwidth]{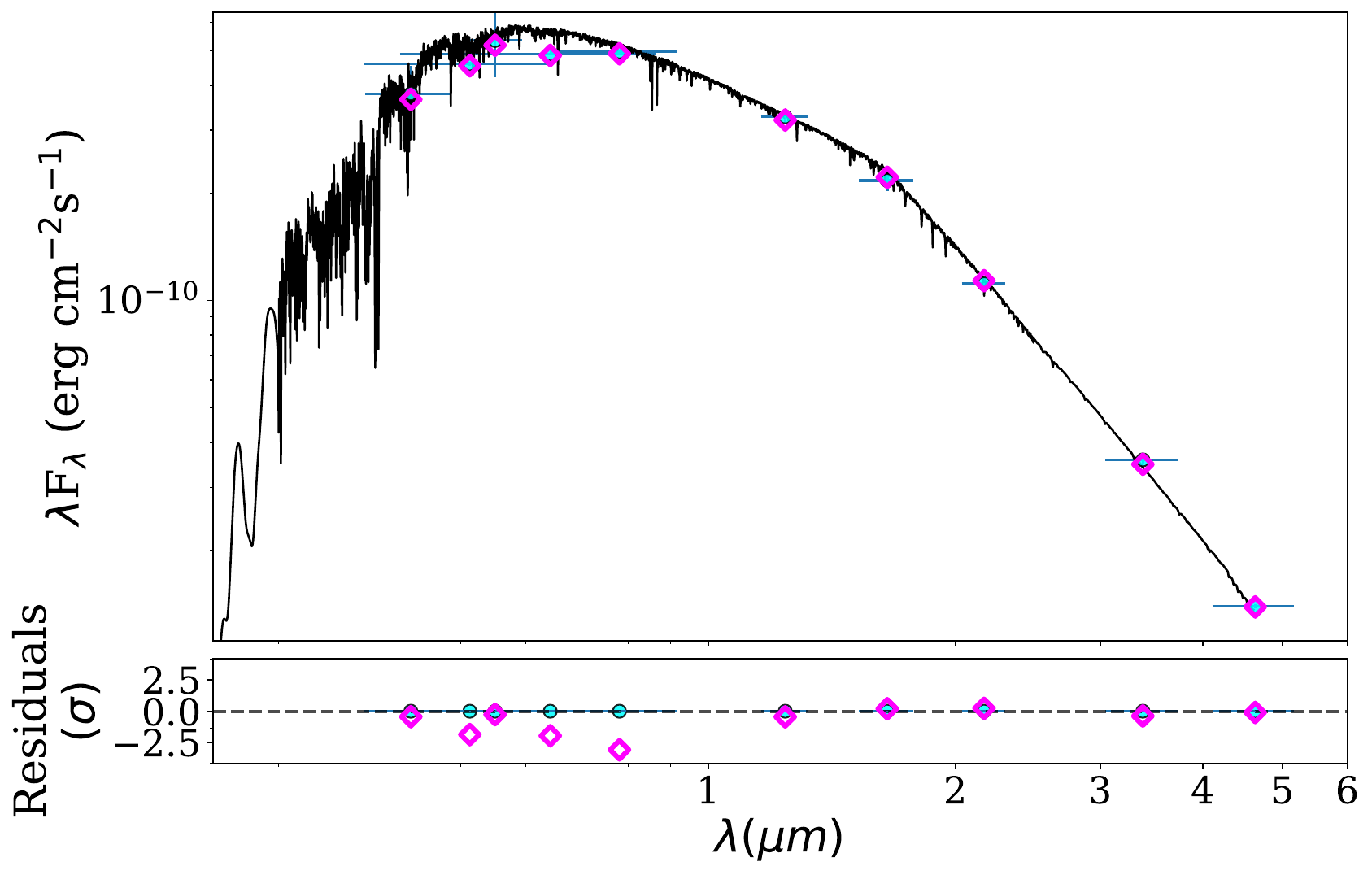}
            \caption{ }
         \label{Figure: SED TOI-2420}
     \end{subfigure}
     \begin{subfigure}[b]{0.49\textwidth}
         \centering
         \includegraphics[width=\textwidth]{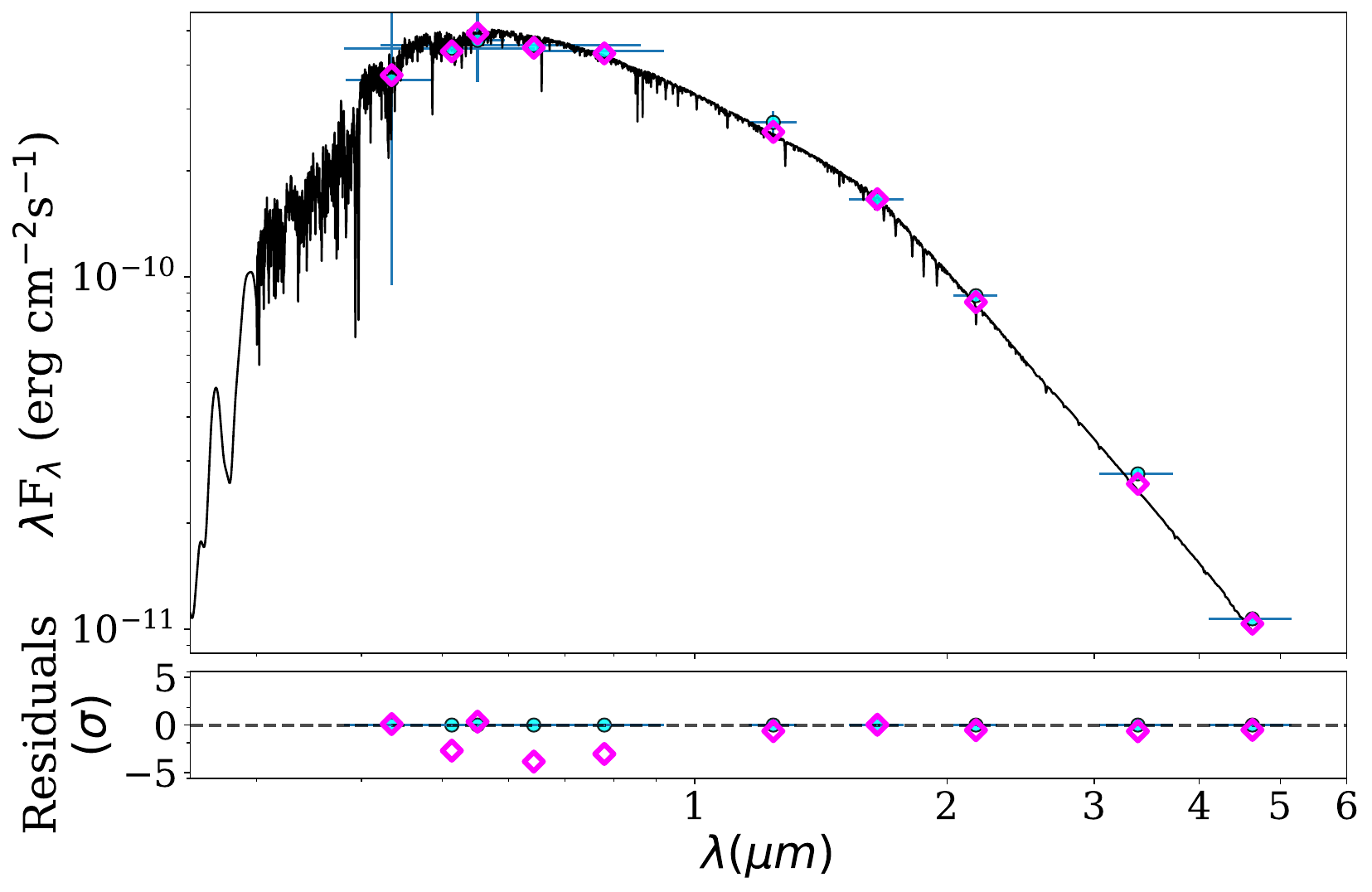}
            \caption{}
        \label{Figure: SED TOI-2485}
     \end{subfigure}
                       \caption{The spectral energy distribution (SED) for TOI-2420 (\emph{left}) and TOI-2485 (\emph{right})  and the best fitted models from 
\citep[{Phoenix~v2}][]{2013A&A...553A...6H}. Magenta and blue diamonds are   the 
synthetic       and the   observed 
photometry, respectively.   One-$\sigma$ uncertainties of the magnitudes are marked with vertical bars,   while the the horizontal bars show the effective width of respective passband. The lower panels shows the residuals  normalised to the errors of the photometry.}       
    \label{Figure: SEDs}
\end{figure*}

The derived spectroscopic parameters from SME were used as priors in  
a spectral energy distribution (SED) fit (Fig. \ref{Figure: SEDs})  with the publically available 
python package   
ARIADNE\footnote{\url{https://github.com/jvines/astroARIADNE}}
\citep[][]{2022MNRAS.513.2719V}. 
This software fits the observed  broadbad photometry 
 to the SED from  grids of four stellar models, constrained by the \textit{Gaia} DR3 parallax and the dust maps of \citet{1998ApJ...500..525S}  to obtain an upper limit on     $A_V$. We included the 
Johnson $V$ 
and $B$  from APASS,
$G G_{\rm BP} G_{\rm RP}$   from \textit{Gaia} DR3,    
$JHK_S$  from {\it 2MASS}, and the {\it WISE} W1 and W2 photometry. The 
atmospheric model grids that were used in the fit were {\tt {Phoenix~v2}} 
\citep{2013A&A...553A...6H}, {\tt {BtSettl}} 
\citep{2012RSPTA.370.2765A}, 
\citet{Castelli2004}, and 
\citet{1993yCat.6039....0K}. The final stellar parameters were computed with Bayesian Model Averaging from the averaged posterior distributions of all four stellar models weighted by respective Bayesian evidence estimate. To account for an underestimation of the uncertainties, an excess noise term is added in ARIADNE to each set of parameters. 

The stellar mass is computed in two ways in ARIADNE. The first method determines a gravitational mass from a combination of the posterior \logg, and the computed \rstar. The second technique used by ARIADNE is an interpolation from   the MIST
\citep{2016ApJ...823..102C} isochrones. 
We note that 
the posteriors of \teff, \logg, and \feh~in the ARIADNE model  are in   good agreement with results   from SME for both targets (listed in Table~\ref{Table: stellar spectroscopic parameters}).

The resulting stellar masses and radii were checked with  the online applet  {\tt {PARAM1.3}}\footnote{\url{http://stev.oapd.inaf.it/cgi-bin/param_1.3}} 
\citep{daSilva2006} based on   
Bayesian computation and the PARSEC isochrones. Input was 
the \textit{Gaia} DR3 parallax, \teff, \feh, the $V$ magnitude. 
The results are in good agreement, within 1~$\sigma$, with the ARIADNE models for both host stars. 

All results are listed in Table~\ref{Table stellar mass and radius} including the  
luminosity and stellar age  derived with ARIADNE and {\tt {PARAM1.3}}. For the modelling of the planets in Sect.~\ref{sec:planet}, we use the ARIADNE results.

\begin{table}[!htb]
   \caption[]{Stellar properties of TOI-2420 and TOI-2485}
     \label{t:star_param}
     \small
     \centering
       \begin{tabular}{lccc}
         \hline
          \hline
         \noalign{\smallskip}
         Parameter   &  \object{TOI-2420}  & \object{TOI-2485} &  Ref \\
         \noalign{\smallskip}
         \hline
         \noalign{\smallskip}
$\alpha$ (J2000)          &    00:59:18.44   &  13:40:49.04      & \textit{Gaia} DR3$^1$    \\
$\delta$ (J2000)          &  -19:46:16.19   & +22:59:02.29 & \textit{Gaia} DR3  \\
$\mu_{\alpha}$ (mas/yr)  &    45.023$\pm$0.033 & 1.024$\pm$0.024
  & \textit{Gaia} DR3  \\
$\mu_{\delta}$ (mas/yr)  &    18.561$\pm$0.034 &  -7.067$\pm$0.015 & \textit{Gaia} DR3  \\
RV     (km\,s$^{-1}$)            &    17.74$\pm$0.43 & -25.81$\pm$0.59  & \textit{Gaia} DR3   \\
$\pi$  (mas)             &    2.249$\pm$0.029 & 2.516$\pm$0.023 & \textit{Gaia} DR3  \\
\noalign{\medskip}
$B$ (mag)                &    12.136$\pm$0.029  & 12.092$\pm$0.312 & APASS DR9$^2$ \\
$V$ (mag)                  &    11.574$\pm$0.092   & 11.935$\pm$0.026  & APASS DR9    \\
$G$ (mag)                  &    11.2863$\pm$0.0007 & 11.3730$\pm$0.0007 & \textit{Gaia} DR3  \\
TESS (mag)              &   10.829$\pm$0.007 &   10.969$\pm$0.008 &  \\
J$_{\rm 2MASS}$ (mag)    &   10.182$\pm$0.023  &  10.371$\pm$0.022& 2MASS$^3$  \\
H$_{\rm 2MASS}$ (mag)    &   9.843$\pm$0.025 & 10.134$\pm$0.030 & 2MASS  \\
K$_{\rm 2MASS}$ (mag)    &   9.800$\pm$0.025 & 10.051$\pm$0.021 & 2MASS  \\
\noalign{\medskip}
         \noalign{\smallskip}
         \hline
        \noalign{\smallskip}
      \end{tabular}
$^1$ \cite{GaiaDR3}, $^2$ \cite{ApassDR9}, $^3$ \cite{Cutri2003}
\end{table}

\begin{table*}
\centering
 \caption{Spectroscopic  parameters for TOI-2420 and TOI-2485  modelled with SME. Posteriors from the ARIADNE modelling and 
the effective stellar temperature from \textit{Gaia} DR2 are listed for comparison.}   
\begin{tabular}{llccccc }
\hline \hline \noalign{\smallskip} \noalign{\smallskip}
    \multicolumn{6}{c}{TOI-2420} \\ \noalign{\smallskip}  
 \hline
     \noalign{\smallskip} \noalign{\smallskip}
Method  & $T_\mathrm{eff}$  & $\log g_\star$ & [Fe/H]   & [Ca/H]& [Na/H] &   $V \sin i_\star$ \\  
& (K)  &(cgs)& (dex) & (dex)& (dex)& (km~s$^{-1}$)    \\
\noalign{\smallskip}
     \hline
\noalign{\smallskip} 
SME$^a$  &  $5537 \pm 70$  & $3.74\pm 0.10$ & $-0.18\pm0.06$   & $-0.10\pm0.04$& $-0.04\pm0.05$ &  $4.1 \pm 0.5$   \\

astroARIADNE$^b$   &  $5560\pm 20 $  & $3.77 \pm 0.08 $ & $-0.19\pm 0.04$   &\ldots  &\ldots&\ldots   \\ 


    \textit{Gaia} DR2 & $5496_{-112}^{+260}$  &\ldots   &\ldots  &\ldots &\ldots &\ldots  \\ \noalign{\smallskip} 
\hline 
  \noalign{\smallskip} \noalign{\smallskip}
    \multicolumn{6}{c}{TOI-2485} \\ \noalign{\smallskip}
 \hline
     \noalign{\smallskip} \noalign{\smallskip}
SME$^a$  &  $5929 \pm 85$  & $4.04\pm 0.07$ & $0.12\pm0.04$   & $0.17\pm0.04$& $ 0.26\pm0.03$ &  $5.5 \pm 0.5$ \\

astroARIADNE$^b$   &  $5939\pm 32 $  & $4.05 \pm 0.09 $ & $0.10\pm 0.04$   &\ldots  &\ldots&\ldots\\ 


    \textit{Gaia} DR2 & $5900_{-41}^{+34}$  &\ldots   &\ldots  &\ldots &\ldots &\ldots\\ \noalign{\smallskip}
\hline 
\end{tabular}  
\label{Table: stellar spectroscopic parameters}
\tablefoot{
\tablefoottext{a}{Adopted as priors for the stellar mass and radius modelling with ARIADNE and PARAM~1.3 in   Sect.~\ref{Subsection: stellar mass and radius}.}
\tablefoottext{b}{Posteriors from Bayesian Model Averinging with ARIADNE.}
}
\end{table*}

  \begin{table*}
 \centering
 \caption{Stellar parameters of  TOI-2420 and TOI-2485 modelled with ARIADNE and PARAM~1.3.}
  \label{Table stellar mass and radius}
  \begin{tabular}{lccccc c}
 \hline\hline
     \noalign{\smallskip} \noalign{\smallskip}
     \multicolumn{6}{c}{TOI-2420} \\ \noalign{\smallskip}
 \hline
     \noalign{\smallskip}
Method    & $M_\star$  &     $R_\star$   & $\rho_\star$ &    $L_\star$ & Age     \\
  & ($M_{\odot}$)  & ($R_{\odot}$)  & (g~cm$^{-3}$) & ($L_{\odot}$) &  (Gyr) \\
    \noalign{\smallskip}
     \hline
\noalign{\smallskip} 
{\tt {astroARIADNE}}$\tablefootmark{a}$    &  $1.158\pm 0.098$  &  $2.369\pm 0.124$ &  $0.12 \pm 0.02$   & $ 4.86\pm 0.51 $ &  $5.3\pm 1.6$  \\
Gravitational mass$\tablefootmark{b}$     & $1.185\pm 0.265$ &  \ldots & \ldots \ldots & \ldots    & \ldots   \\

PARAM 1.3   & $1.206\pm0.034$  &  $2.277\pm0.092$ &  $0.14\pm 0.02$ & \ldots& $4.6\pm 0.4$\\ 

\textit{Gaia} DR2  & \ldots    &   $2.345^{+0.098}_{-0.207}$ & \ldots  & \ldots  & \ldots \\    \noalign{\smallskip}

 \hline 
  \noalign{\smallskip} \noalign{\smallskip}
    \multicolumn{6}{c}{TOI-2485} \\ \noalign{\smallskip}
 \hline
     \noalign{\smallskip} \noalign{\smallskip}
{\tt {astroARIADNE}}$\tablefootmark{a}$    &  $1.163\pm0.053$  &  $1.720\pm 0.069$ &  $0.32 \pm 0.04$   & $ 3.31\pm 0.28 $ &  $6.0^{+0.8}_{-1.7}$  \\
Gravitational mass$\tablefootmark{b}$     & $1.167\pm 0.127$ &  \ldots & \ldots \ldots & \ldots    & \ldots   \\

PARAM 1.3   & $1.210\pm0.048$  &  $1.625\pm0.047$ &  $0.40\pm 0.04$ & \ldots& $4.7\pm 0.8$\\ 

\textit{Gaia} DR2  & \ldots    &   $1.760^{+0.030}_{-0.020}$ & \ldots  & \ldots  & \ldots \\   
     
      \noalign{\smallskip} \noalign{\smallskip}
\hline 
\end{tabular}
\tablefoot{
\tablefoottext{a}{ARIANDE uses SED fitting and MIST isochrones. We adopt these results as the final stellar parameters in the joint transit and RV modelling in Sect.~\ref{sec:planet}.} 
\tablefoottext{b}{Gravitational mass computed from  \logg~and \rstar~modelled with {\tt {astroARIADNE}}.} 
}
\end{table*}


\section{Planetary system modelling: joint fit} 
\label{sec:planet}

We performed a joint RV and transit modelling for \toioneb\  and \toitwob.
We use the code \pyaneti\, \citep{pyaneti,pyaneti2} to model all of our data.

For the transit analyses, we use the quadratic limb darkening framework by \citet{Mandel2002}. We use the $q_1$ and $q_2$ parametrisation given by \citet{Kipping2013} to account for realistic limb darkening parameter values. 
We note that the FFI data are taken with long cadence of 30 and 10 min in different  TESS sectors. For these cases we re-sampled the model to account for the data integration \citep{Kipping2010}, using one integration step for every minute of integration of the data.
For each planet we sample for the time of transit, $T_0$; orbital period, $P_{\rm orb}$; the polar parametrisation of the orbital eccentricity, $e$ and angle of periastron, $\omega$ given as $\sqrt{e} \cos \omega$ and $\sqrt{e} \sin \omega$ \citep[see][]{Anderson2011}; scaled planetary radius $R_{\rm p}/R_\star$ and stellar density $\rho_\star$ (that connects with the scaled semi-major axis $a/R_\star$ via Kepler's third law). 
We also sample for a photometric jitter term per band to penalise the imperfections of our transit model.

For the RV data, we use one Keplerian signal for each system. This Keplerian signal is modelled with a time of minimum conjunction (or time of transit for transiting planets), $T_0$; orbital period, $P_{\rm orb}$; orbital eccentricity, $\sqrt{e} \cos \omega$ and $\sqrt{e} \sin \omega$; and Doppler semi-amplitude, $K$. 
We also include one offset to account for the systemic offset and a jitter term for every instrument in the corresponding data set.
For TOI-2485, we also included a slope to model the trend visible in the FEROS time series.

Tables~\ref{tab:parstoi2420} and \ref{tab:parstoi2485} show the sampled parameters and priors used to model \toioneb\ and \toitwob, respectively.
In all our runs we sample the parameter space with 250 walkers using a Markov chain Monte Carlo (MCMC) ensemble sampler algorithm \citep[as implemented in \texttt{pyaneti}][]{pyaneti,emcee}. 
We create the posterior distributions using the last 5000 iterations of converged chains, thinned with a thin factor of 10. This gives a distribution of 125\,000  points for each sampled parameter. 

We ran different model combinations to model \toioneb\ and \toitwob, including circular and eccentric orbits, and linear and quadratic trends. We use the difference of Akaike Information Criterion ($\Delta$\,AIC) to find the best model. We decide to use the AIC because is more appropriate than BIC (Bayesian Information Criterion) in finding the best model when the true model is unknown \citep[see discussion in][]{Barragan2023}. Table~\ref{tab:aics} summarises the results. We can conclude that the best model for TOI-2420 is an eccentric orbit with no trends on the RVs. As for TOI-2485 the best model is the one with a quadratic trend and an eccentric orbit. We also tested a 2-planet model, but the fit does not converge to any significant results.

\begin{figure}
    \centering
    \includegraphics[width=0.48\textwidth]{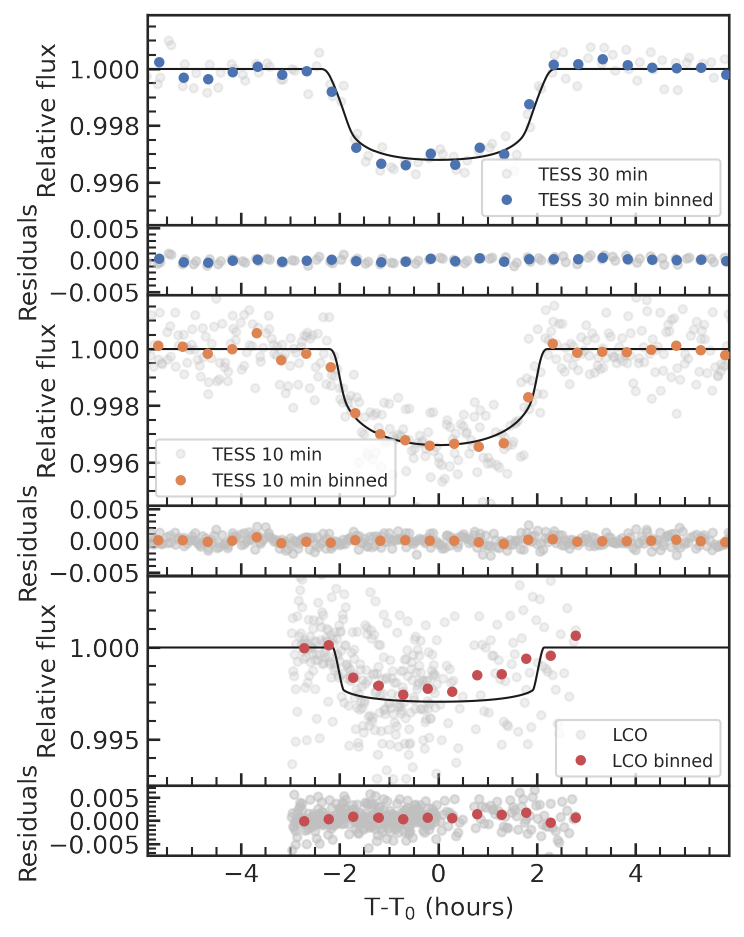}\\
    \includegraphics[width=0.48\textwidth]{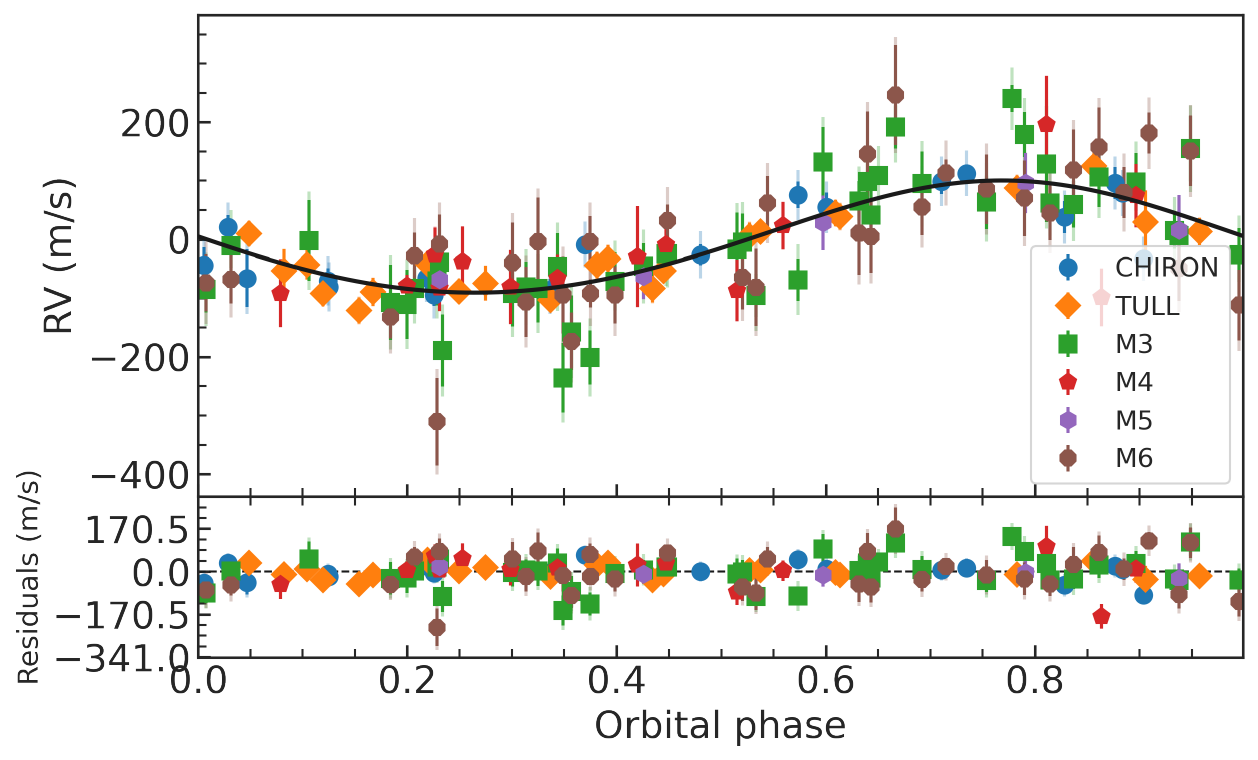}
    \caption{\emph{Top panel:} Phase-folded transit light curve \toioneb. Nominal TESS and LCO observations are shown in light grey. Solid coloured circles represent the binned data. Transit models are shown with a solid black line. \emph{Bottom panel:} Phase-folded RV signal for \toioneb\ following the subtraction of the systemic velocity. Blue circles and triangles show the CHIRON and TULL RV data, respectively, while green squares, red pentagons, purple hexagons and brown circles show the \textsc{Minerva} RVs, split in 4 different datasets.}
    \label{fig:toi2420fits}
\end{figure}

\begin{figure}
    \centering
    \includegraphics[width=0.48\textwidth]{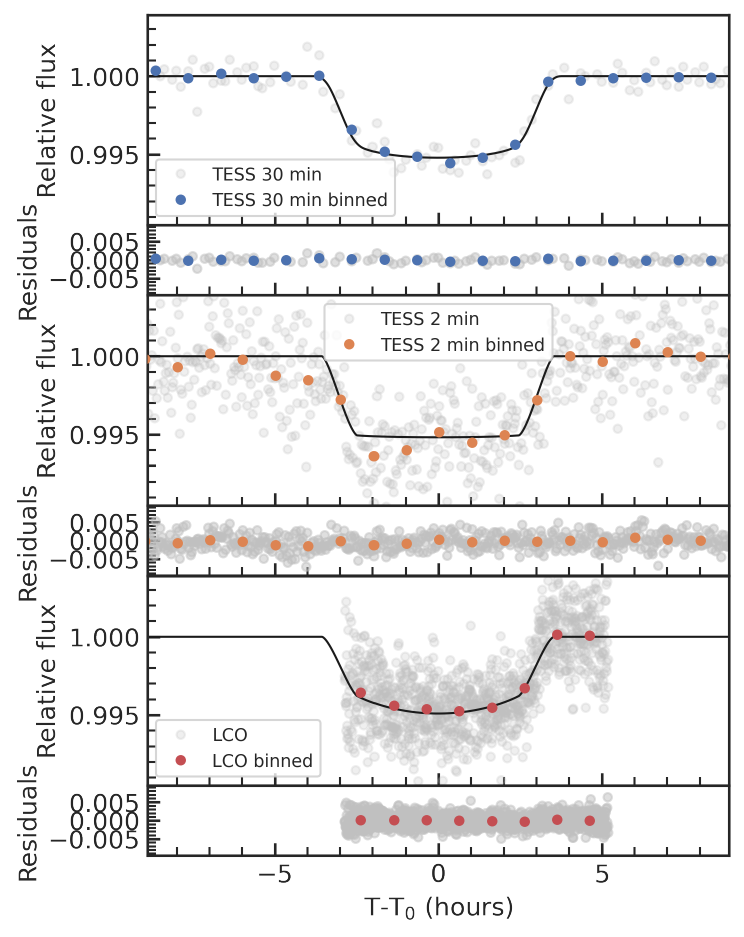}\\
    \includegraphics[width=0.48\textwidth]{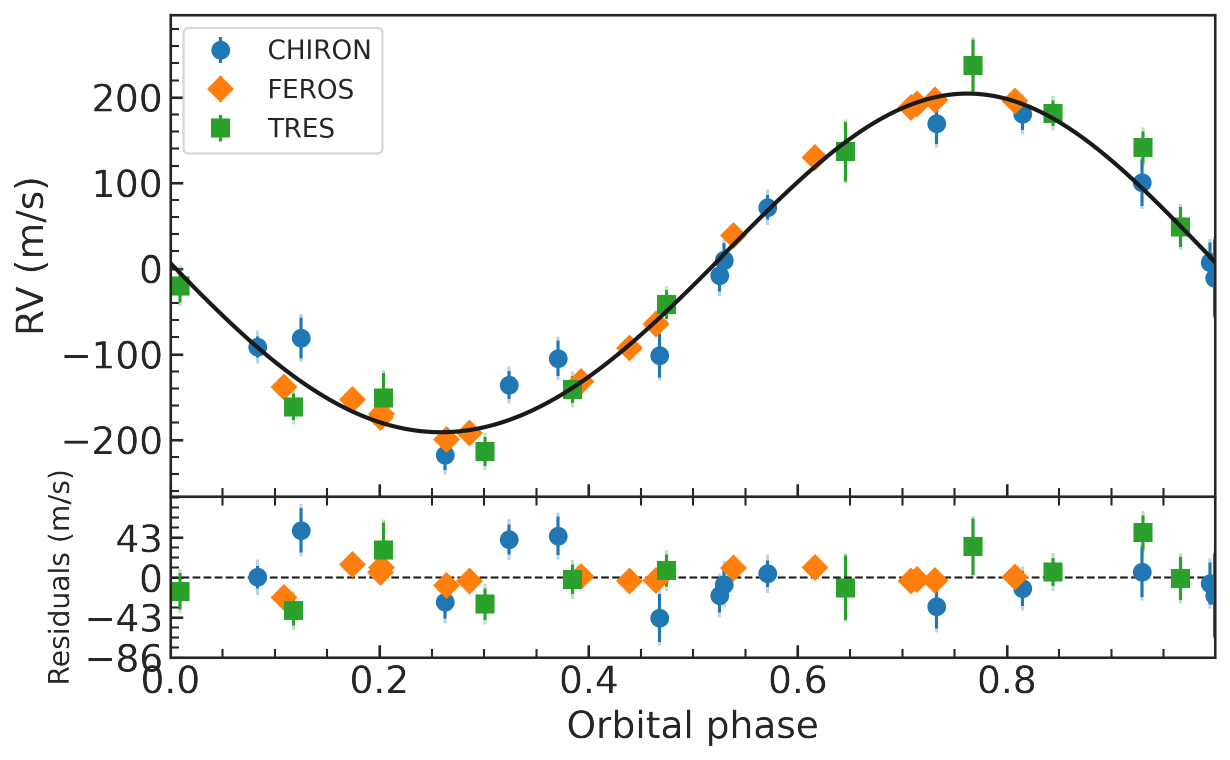}
    \caption{\emph{Top panel:} Phase-folded transit light curve \toitwob. Nominal TESS and LCO observations are shown in light grey. Solid coloured circles represent the binned data. Transit models are shown with a solid black line. \emph{Bottom panel:} Phase-folded RV signal for \toitwob\ following the subtraction of the systemic velocities. Orange circles, diamonds and squares show CHIRON, FEROS and TRES RV data, respectively. }
    \label{fig:toi2485fits}
\end{figure}

\begin{figure*}
    \centering
    \includegraphics[width=0.99\textwidth]{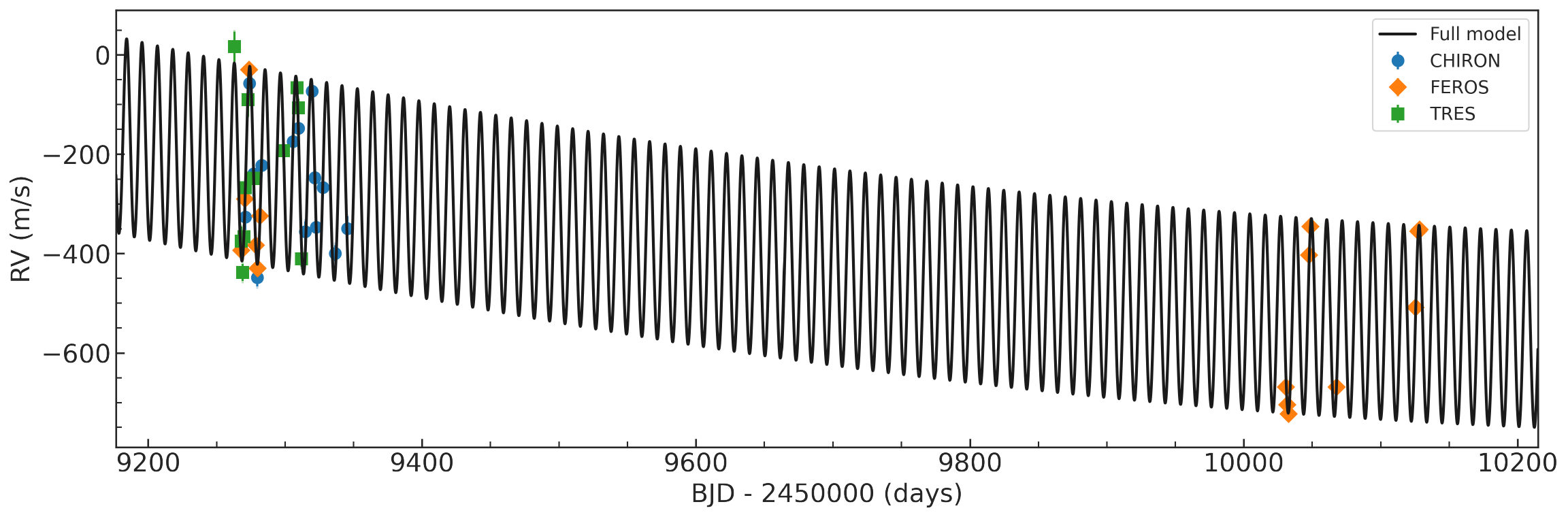}
    \caption{TOI-2485's RV time series. The trend in the data is clear and it has been modelled with a quadratic trend, whose significance is higher with respect of a} linear trend model.
    \label{fig:toi2485rvtimeseries}
\end{figure*}

The inferred and derived parameters of \toioneb\ and \toitwob\ are shown in Tables~\ref{tab:parstoi2420} and \ref{tab:parstoi2485}. Figures~\ref{fig:toi2420fits} and \ref{fig:toi2485fits} show the inferred transit and RV models for both planets, while Fig. \ref{fig:toi2485rvtimeseries} displays the RVs time series for TOI-2485 where a linear trend is evident. We note that the TESS 2 min data in Fig. \ref{fig:toi2485fits} looks flat-bottomed. However, we underline that we account for the limb darkening coefficients in the modelling, using uniform. In this particular case, the best solution is consistent with a flat bottom transit, suggesting that we cannot constrain the limb darkening of the star on the TESS band with that given transit dataset. However, we note that this does not affect the inferred transit depth.

\begin{table*}
\begin{center}
\caption{Model comparison for different models for TOI-2420 and TOI-2485. Each element in the table shows the $\Delta \mathrm{AIC}$ for each model in comparison with the minimum AIC value for each system. \label{tab:aics}} 
\begin{tabular}{rcccc}
\hline\hline
Model  &  \multicolumn{2}{c}{\emph{ \bf TOI-2420}} & \multicolumn{2}{c}{\emph{ \bf TOI-2485}}  \\
& Circular & Eccentric & Circular & Eccentric   \\
\hline
No trend & 2 & \textbf{0} & $\cdots$ &  $\cdots$ \\
Linear trend & 3 & 2 & 12 & 5 \\
Quadratic trend & $\cdots$ &  $\cdots$ & 12 & \textbf{0} \\
\hline
\end{tabular}
\end{center}
\end{table*}

\newcommand{\smass}[1][$M_{\odot}$]{ $ 1.1580000 _{- 0.0980000}^{ + 0.0980000} $ #1} 
\newcommand{\sradius}[1][$R_{\odot}$]{ $2.3690000 _{ - 0.1240000}^{ + 0.1240000} $ #1}
\newcommand{\stemp}[1][$\mathrm{K}$]{ $ 5537.0000000 _{- 70.0000000}^{ + 70.0000000} $ #1 }
\newcommand{\Tzerob}[1][days]{ $ 8388.4119 \pm 0.0013 $~#1 } 
\newcommand{\Pb}[1][days]{ $ 5.842641_{-0.000013}^{+0.000015} $~#1 } 
\newcommand{\esinb}[1][ ]{ $ 0.05_{-0.20}^{+0.16} $~#1 } 
\newcommand{\ecosb}[1][ ]{ $ 0.159_{-0.104}^{+0.074} $~#1 } 
\newcommand{\bb}[1][ ]{ $ 0.849_{-0.025}^{+0.018} $~#1 } 
\newcommand{\dentrheeb}[1][${\rm g^{1/3}\,cm^{-1}}$]{ $ 0.133_{-0.026}^{+0.036} $~#1 } 
\newcommand{\rrb}[1][ ]{ $ 0.05810_{-0.00083}^{+0.00098} $~#1 } 
\newcommand{\kb}[1][${\rm m\,s^{-1}}$]{ $ 94.25_{-6.29}^{+6.58} $~#1 } 
\newcommand{\mpb}[1][$M_\mathrm{J}$]{ $ 0.927_{-0.079}^{+0.085} $~#1 } 
\newcommand{\rpb}[1][$R_\mathrm{J}$]{ $ 1.340_{-0.072}^{+0.074} $~#1 } 
\newcommand{\Tperib}[1][days]{ $ 8387.35_{-1.05}^{+0.74} $~#1 } 
\newcommand{\eb}[1][ ]{ $ 0.055_{-0.031}^{+0.036} $~#1 } 
\newcommand{\wb}[1][deg]{ $ 15.5_{-64.4}^{+47.5} $~#1 } 
\newcommand{\prvb}[1][${\rm km\,s^{-1}}$]{ $ nan_{-nan}^{+nan} $~#1 } 
\newcommand{\ib}[1][deg]{ $ 82.07_{-1.04}^{+0.98} $~#1 } 
\newcommand{\arb}[1][ ]{ $ 6.21_{-0.43}^{+0.51} $~#1 } 
\newcommand{\ab}[1][AU]{ $ 0.0684_{-0.0059}^{+0.0067} $~#1 } 
\newcommand{\insolationb}[1][${\rm F_{\oplus}}$]{ $ 1017_{-156}^{+167} $~#1 } 
\newcommand{\tsmb}[1][ ]{ $ 13.78_{-1.53}^{+1.76} $~#1 } 
\newcommand{\denstrb}[1][${\rm g\,cm^{-3}}$]{ $ 0.133_{-0.026}^{+0.036} $~#1 } 
\newcommand{\densspb}[1][${\rm g\,cm^{-3}}$]{ $ 0.123_{-0.020}^{+0.024} $~#1 } 
\newcommand{\Teqb}[1][K]{ $ 1571.6_{-64.0}^{+60.8} $~#1 } 
\newcommand{\ttotb}[1][hours]{ $ 4.554_{-0.092}^{+0.093} $~#1 } 
\newcommand{\tfulb}[1][hours]{ $ 2.93_{-0.15}^{+0.16} $~#1 } 
\newcommand{\tegb}[1][hours]{ $ 0.81 \pm 0.11 $~#1 } 
\newcommand{\deltamagb}[1][]{ $ 2.85_{-0.33}^{+0.28} $~#1 } 
\newcommand{\denpb}[1][${\rm g\,cm^{-3}}$]{ $ 0.477_{-0.080}^{+0.099} $~#1 } 
\newcommand{\grapb}[1][${\rm cm\,s^{-2}}$]{ $ 1346_{-193}^{+251} $~#1 } 
\newcommand{\grapparsb}[1][${\rm cm\,s^{-2}}$]{ $ 1283_{-167}^{+193} $~#1 } 
\newcommand{\jspb}[1][ ]{ $ 94.78_{-9.95}^{+10.91} $~#1 } 
\newcommand{\qoneSL}[1][]{ $ 0.096_{-0.063}^{+0.123} $~#1 } 
\newcommand{\qtwoSL}[1][]{ $ 0.34_{-0.23}^{+0.35} $~#1 } 
\newcommand{\uoneSL}[1][]{ $ 0.19_{-0.14}^{+0.20} $~#1 } 
\newcommand{\utwoSL}[1][]{ $ 0.09_{-0.17}^{+0.21} $~#1 } 
\newcommand{\qoneSC}[1][]{ $ 0.089_{-0.062}^{+0.145} $~#1 } 
\newcommand{\qtwoSC}[1][]{ $ 0.22_{-0.16}^{+0.23} $~#1 } 
\newcommand{\uoneSC}[1][]{ $ 0.119_{-0.088}^{+0.153} $~#1 } 
\newcommand{\utwoSC}[1][]{ $ 0.15_{-0.13}^{+0.19} $~#1 } 
\newcommand{\qonegblc}[1][]{ $ 0.80_{-0.18}^{+0.14} $~#1 } 
\newcommand{\qtwogblc}[1][]{ $ 0.470_{-0.149}^{+0.096} $~#1 } 
\newcommand{\uonegblc}[1][]{ $ 0.83_{-0.26}^{+0.12} $~#1 } 
\newcommand{\utwogblc}[1][]{ $ 0.05_{-0.16}^{+0.27} $~#1 } 
\newcommand{\CHIRON}[1][${\rm km\,s^{-1}}$]{ $ 15.8940_{-0.0097}^{+0.0108} $~#1 } 
\newcommand{\TULL}[1][${\rm km\,s^{-1}}$]{ $ 10.2195_{-0.0059}^{+0.0060} $~#1 } 
\newcommand{\Mthree}[1][${\rm km\,s^{-1}}$]{ $ 16.5141_{-0.0094}^{+0.0092} $~#1 } 
\newcommand{\Mfour}[1][${\rm km\,s^{-1}}$]{ $ 16.502_{-0.017}^{+0.015} $~#1 } 
\newcommand{\Mfive}[1][${\rm km\,s^{-1}}$]{ $ 16.597 \pm 0.023 $~#1 } 
\newcommand{\Msix}[1][${\rm km\,s^{-1}}$]{ $ 16.552_{-0.013}^{+0.012} $~#1 } 
\newcommand{\jCHIRON}[1][${\rm m\,s^{-1}}$]{ $ 35.96_{-7.31}^{+10.31} $~#1 } 
\newcommand{\jTULL}[1][${\rm m\,s^{-1}}$]{ $ 10.06_{-8.17}^{+10.39} $~#1 } 
\newcommand{\jMthree}[1][${\rm m\,s^{-1}}$]{ $ 50.72_{-8.74}^{+10.42} $~#1 } 
\newcommand{\jMfour}[1][${\rm m\,s^{-1}}$]{ $ 22.5_{-20.3}^{+28.5} $~#1 } 
\newcommand{\jMfive}[1][${\rm m\,s^{-1}}$]{ $ 5.76_{-4.96}^{+26.22} $~#1 } 
\newcommand{\jMsix}[1][${\rm m\,s^{-1}}$]{ $ 50.0_{-13.1}^{+13.4} $~#1 } 
\newcommand{\jtrSL}[1][ppm]{ $ 398.2_{-22.1}^{+27.5} $~#1 } 
\newcommand{\jtrSC}[1][ppm]{ $ 858.8_{-25.1}^{+26.8} $~#1 } 
\newcommand{\jtrgblc}[1][ppm]{ $ 2384.0_{-89.8}^{+91.5} $~#1 } 

\begin{table*}
\begin{center}
  \caption{Model parameters and priors for \toioneb's joint fit \label{tab:parstoi2420}}  
  \begin{tabular}{lcc}
  \hline
  \hline
  \noalign{\smallskip}
  Parameter & Prior\tablefootmark{a} & Inferred paramter\tablefootmark{b} \\
  \noalign{\smallskip}
  \hline
  \noalign{\smallskip}
  \multicolumn{3}{l}{\emph{\bf \toioneb's sampled parameters }} \\
  \noalign{\smallskip}
    Orbital period $P_{\mathrm{orb}}$ (days)  & $\mathcal{U}[5.840 , 5.845 ]$ &\Pb[] \\
    Transit epoch $T_0$ (BJD$_\mathrm{TDB}-$2\,450\,000)  & $\mathcal{U}[8388.35 , 8388.45]$ & \Tzerob[]  \\ 
  $e$ and $\omega$ polar parametrisation, $\sqrt{e} \cos \omega$ & $\mathcal{U}[-1,1]$  & \ecosb \\
  \noalign{\smallskip}
  $e$ and $\omega$ polar parametrisation, $\sqrt{e} \sin \omega$ & $\mathcal{U}[-1,1]$  & \esinb \\
   \noalign{\smallskip}
   Scaled planet radius $R_\mathrm{p}/R_{\star}$  &$\mathcal{U}[0.0,0.2]$ & \rrb[]  \\
  \noalign{\smallskip}
    Impact parameter, $b$ & $\mathcal{U}[0,1]$ & \bb[] \\
  \noalign{\smallskip}
    Stellar density $\rho_\star$ (${\rm g\,cm^{-3}}$) & $\mathcal{U}[0.01,1]$ & \denstrb[]   \\
  \noalign{\smallskip}
    Doppler semi-amplitude variation $K$ (m s$^{-1}$) & $\mathcal{U}[0,500]$ & \kb[] \\
    \multicolumn{3}{l}{\emph{ \bf Other sampled parameters}} \\
    Offset CHIRON (\kms) & $\mathcal{U}[ 15.31 , 16.51]$ & \CHIRON[] \\
  \noalign{\smallskip}
    Offset TULL (\kms) & $\mathcal{U}[ 9.31 , 10.51]$ & \TULL[] \\
  \noalign{\smallskip}
    Offset \textsc{Minerva} 3 (\kms) & $\mathcal{U}[ 15.78 , 17.26 ]$ & \Mthree[] \\
  \noalign{\smallskip}
   Offset \textsc{Minerva} 4 (\kms) & $\mathcal{U}[ 15.78 , 17.26 ]$ & \Mfour[] \\
    Offset \textsc{Minerva} 5 (\kms) & $\mathcal{U}[ 15.78 , 17.26 ]$ & \Mfive[] \\
    Offset \textsc{Minerva} 6 (\kms) & $\mathcal{U}[ 15.78 , 17.26 ]$ & \Msix[] \\
  \noalign{\smallskip}
    Jitter term $\sigma_{\rm CHIRON}$ (\ms) & $\mathcal{J}[1,100]$ & \jCHIRON[] \\
  \noalign{\smallskip}
    Jitter term $\sigma_{\rm TULL}$ (\ms) & $\mathcal{J}[1,100]$ & \jTULL[] \\
  \noalign{\smallskip}
    Jitter term $\sigma_{\rm M3}$ (\ms) & $\mathcal{J}[1,100]$ & \jMthree[] \\
  \noalign{\smallskip}
    Jitter term $\sigma_{\rm M4}$ (\ms) & $\mathcal{J}[1,100]$ & \jMfour[] \\
  \noalign{\smallskip}
    Jitter term $\sigma_{\rm M5}$ (\ms) & $\mathcal{J}[1,100]$ & \jMfive[] \\
  \noalign{\smallskip}
    Jitter term $\sigma_{\rm M6}$ (\ms) & $\mathcal{J}[1,100]$ & \jMsix[] \\
  \noalign{\smallskip}
    TESS S03 limb-darkening coefficient $q_1$  &$\mathcal{U}[0,1]$ & \qoneSL \\ 
  \noalign{\smallskip}
    TESS S03 limb-darkening coefficient $q_2$  &$\mathcal{U}[0,1]$ & \qtwoSL \\ 
  \noalign{\smallskip}
    TESS S30 limb-darkening coefficient $q_1$  &$\mathcal{U}[0,1]$ & \qoneSC \\ 
  \noalign{\smallskip}
    TESS S30 limb-darkening coefficient $q_2$  &$\mathcal{U}[0,1]$ & \qtwoSC \\ 
  \noalign{\smallskip}
    LCO limb-darkening coefficient $q_1$  &$\mathcal{U}[0,1]$ & \qonegblc \\ 
  \noalign{\smallskip}
    LCO limb-darkening coefficient $q_2$  &$\mathcal{U}[0,1]$ & \qtwogblc \\ 
  \noalign{\smallskip}
    Jitter term $\sigma_{\rm TESS,S03}$ (ppm) & $\mathcal{J}[1,100]$ & \jtrSL[] \\
  \noalign{\smallskip}
    Jitter term $\sigma_{\rm TESS,S30}$ (ppm) & $\mathcal{J}[1,100]$ & \jtrSC[] \\
  \noalign{\smallskip}
    Jitter term $\sigma_{\rm LCO}$ (ppm) & $\mathcal{J}[1,100]$ & \jtrgblc[] \\
    \multicolumn{3}{l}{\emph{ \bf \toioneb's derived parameters}} \\
    Planet mass $M_\mathrm{p}$ ($M_{\rm J}$) & $\cdots$ &  \mpb[]     \\
  \noalign{\smallskip}
    Planet radius $R_\mathrm{p}$ ($R_{\rm J}$) & $\cdots$ &  \rpb[]   \\
  \noalign{\smallskip}
    Planet density $\rho_{\rm p}$ (g\,cm$^{-3}$) & $\cdots$ &  \denpb[]    \\
  \noalign{\smallskip}
    Orbital eccentricity, $e$ & $\cdots$ & \eb[] \\
  \noalign{\smallskip}
    Angle of periastron, $\omega$(deg) & $\cdots$ & \wb[] \\
  \noalign{\smallskip}
    Scaled semi-major axis  $a/R_\star$ & $\cdots$ & \arb[]   \\
  \noalign{\smallskip}
    Semi-major axis  $a$ (AU) & $\cdots$ &  \ab[]  \\
    Time of periastron passage $T_p$ (BJD-2450000) & $\cdots$ & \Tzerob[]  \\
    Orbit inclination $i_\mathrm{p}$ ($^{\circ}$)  & $\cdots$ & \ib[]    \\
  \noalign{\smallskip}
    Total transit duration $\tau_{14}$ (hours) & $\cdots$ & \ttotb[]   \\
  \noalign{\smallskip}
    Planet surface gravity $g_{\rm p}$ (${\rm cm\,s^{-2}}$) & $\cdots$ & \grapb[] \\
  \noalign{\smallskip}
    Equilibrium temperature  $T_\mathrm{eq}$ (K)\tablefootmark{c} & $\cdots$  &   \Teqb[]  \\
  \noalign{\smallskip}
    Received irradiance ($F_\oplus$) & $\cdots$ & \insolationb[]  \\
    \hline
  \end{tabular}
\tablefoot{
\tablefoottext{a}{$\mathcal{U}[a,b]$ refers to an uniform prior between $a$ and $b$, $\mathcal{N}[a,b]$ to a Gaussian prior with mean $a$ and standard deviation $b$, and $\mathcal{J}[a,b]$ to the modified Jeffrey's prior as defined by \citet[eq.~16]{Gregory2005}.} 
\tablefoottext{b}{Inferred parameters and errors are defined as the median and 68.3\% credible interval of the posterior distribution.} 
\tablefoottext{c}{Assuming a zero albedo.}
}
\end{center}
\end{table*}

\newcommand{\Tzerobtwo}[1][days]{ $ 8939.7856_{-0.0023}^{+0.0022} $~#1 } 
\newcommand{\Pbtwo}[1][days]{ $ 11.234790_{-0.000052}^{+0.000054} $~#1 } 
\newcommand{\esinbtwo}[1][ ]{ $ 0.030_{-0.106}^{+0.091} $~#1 } 
\newcommand{\ecosbtwo}[1][ ]{ $ 0.162_{-0.031}^{+0.023} $~#1 } 
\newcommand{\bbtwo}[1][ ]{ $ 0.121_{-0.087}^{+0.134} $~#1 } 
\newcommand{\dentrheebtwo}[1][${\rm g^{1/3}\,cm^{-1}}$]{ $ 0.385_{-0.036}^{+0.031} $~#1 } 
\newcommand{\rrbtwo}[1][ ]{ $ 0.06470_{-0.00068}^{+0.00067} $~#1 } 
\newcommand{\kbtwo}[1][${\rm m\,s^{-1}}$]{ $ 197.84_{-3.80}^{+3.99} $~#1 } 
\newcommand{\mpbtwo}[1][$M_\mathrm{J}$]{ $ 2.412_{-0.087}^{+0.088} $~#1 } 
\newcommand{\Tperibtwo}[1][days]{ $ 8937.40_{-1.13}^{+0.87} $~#1 } 
\newcommand{\ebtwo}[1][ ]{ $ 0.0341_{-0.0087}^{+0.0109} $~#1 } 
\newcommand{\wbtwo}[1][deg]{ $ 10.0_{-36.0}^{+27.7} $~#1 } 
\newcommand{\prvbtwo}[1][${\rm km\,s^{-1}}$]{ $ nan_{-nan}^{+nan} $~#1 } 
\newcommand{\ibtwo}[1][deg]{ $ 89.49_{-0.60}^{+0.36} $~#1 } 
\newcommand{\arbtwo}[1][ ]{ $ 13.69_{-0.44}^{+0.36} $~#1 } 
\newcommand{\abtwo}[1][AU]{ $ 0.1093_{-0.0057}^{+0.0055} $~#1 } 
\newcommand{\insolationbtwo}[1][${\rm F_{\oplus}}$]{ $ 275.6_{-23.5}^{+27.2} $~#1 } 
\newcommand{\rpbtwo}[1][$R_\mathrm{J}$]{ $ 1.083 \pm 0.045 $~#1 } 
\newcommand{\tsmbtwo}[1][ ]{ $ 3.83_{-0.24}^{+0.27} $~#1 } 
\newcommand{\denstrbtwo}[1][${\rm g\,cm^{-3}}$]{ $ 0.385_{-0.036}^{+0.031} $~#1 } 
\newcommand{\densspbtwo}[1][${\rm g\,cm^{-3}}$]{ $ 0.322_{-0.038}^{+0.045} $~#1 } 
\newcommand{\Teqbtwo}[1][K]{ $ 1134.0_{-25.0}^{+27.0} $~#1 } 
\newcommand{\ttotbtwo}[1][hours]{ $ 6.567 \pm 0.088 $~#1 } 
\newcommand{\tfulbtwo}[1][hours]{ $ 5.743_{-0.087}^{+0.083} $~#1 } 
\newcommand{\tegbtwo}[1][hours]{ $ 0.4057_{-0.0084}^{+0.0223} $~#1 } 
\newcommand{\deltamagbtwo}[1][]{ $ 0.036_{-0.030}^{+0.115} $~#1 } 
\newcommand{\denpbtwo}[1][${\rm g\,cm^{-3}}$]{ $ 2.36_{-0.28}^{+0.33} $~#1 } 
\newcommand{\grapbtwo}[1][${\rm cm\,s^{-2}}$]{ $ 5749_{-390}^{+317} $~#1 } 
\newcommand{\grapparsbtwo}[1][${\rm cm\,s^{-2}}$]{ $ 5112_{-438}^{+489} $~#1 } 
\newcommand{\jspbtwo}[1][ ]{ $ 421.9_{-24.5}^{+25.5} $~#1 } 
\newcommand{\qoneSLtwo}[1][]{ $ 0.69_{-0.26}^{+0.21} $~#1 } 
\newcommand{\qtwoSLtwo}[1][]{ $ 0.32_{-0.17}^{+0.23} $~#1 } 
\newcommand{\uoneSLtwo}[1][]{ $ 0.53_{-0.25}^{+0.23} $~#1 } 
\newcommand{\utwoSLtwo}[1][]{ $ 0.29_{-0.37}^{+0.33} $~#1 } 
\newcommand{\qoneSCtwo}[1][]{ $ 0.73_{-0.28}^{+0.19} $~#1 } 
\newcommand{\qtwoSCtwo}[1][]{ $ 0.18_{-0.12}^{+0.19} $~#1 } 
\newcommand{\uoneSCtwo}[1][]{ $ 0.30_{-0.20}^{+0.26} $~#1 } 
\newcommand{\utwoSCtwo}[1][]{ $ 0.53_{-0.35}^{+0.25} $~#1 } 
\newcommand{\qonegblctwo}[1][]{ $ 0.148_{-0.064}^{+0.124} $~#1 } 
\newcommand{\qtwogblctwo}[1][]{ $ 0.54_{-0.29}^{+0.30} $~#1 } 
\newcommand{\uonegblctwo}[1][]{ $ 0.40_{-0.17}^{+0.14} $~#1 } 
\newcommand{\utwogblctwo}[1][]{ $ -0.03_{-0.17}^{+0.28} $~#1 } 
\newcommand{\CHIRONtwo}[1][${\rm km\,s^{-1}}$]{ $ -27.563_{-0.050}^{+0.046} $~#1 } 
\newcommand{\FEROStwo}[1][${\rm km\,s^{-1}}$]{ $ -26.148_{-0.047}^{+0.042} $~#1 } 
\newcommand{\TREStwo}[1][${\rm km\,s^{-1}}$]{ $ 0.371_{-0.049}^{+0.044} $~#1 } 
\newcommand{\jCHIRONtwo}[1][${\rm m\,s^{-1}}$]{ $ 14.7_{-11.9}^{+11.1} $~#1 } 
\newcommand{\jFEROStwo}[1][${\rm m\,s^{-1}}$]{ $ 3.14_{-2.54}^{+5.09} $~#1 } 
\newcommand{\jTREStwo}[1][${\rm m\,s^{-1}}$]{ $ 13.8_{-11.4}^{+12.5} $~#1 } 
\newcommand{\jtrSLtwo}[1][ppm]{ $ 702.8_{-41.7}^{+50.4} $~#1 } 
\newcommand{\jtrSCtwo}[1][ppm]{ $ 2123.4_{-56.5}^{+59.5} $~#1 } 
\newcommand{\jtrgblctwo}[1][ppm]{ $ 1766.3_{-37.0}^{+39.1} $~#1 } 
\newcommand{\ltrendtwo}[1][${\rm m\,s^{-1}\,d^{-1}}$]{ $ -0.77_{-0.16}^{+0.18} $~#1 } 
\newcommand{\qtrendtwo}[1][${\rm m\,s^{-1}\,d^{-2}}$]{ $ 0.26_{-0.12}^{+0.11} $~#1 } 

\begin{table*}
\begin{center}
  \caption{Model parameters and priors for \toitwob's joint fit \label{tab:parstoi2485}}  
  \begin{tabular}{lcc}
  \hline
  \hline
  \noalign{\smallskip}
  Parameter & Prior\tablefootmark{a} & Inferred parameter\tablefootmark{b} \\
  \noalign{\smallskip}
  \hline
  \noalign{\smallskip}
  \multicolumn{3}{l}{\emph{\bf \toitwob's sampled parameters }} \\
  \noalign{\smallskip}
    Orbital period $P_{\mathrm{orb}}$ (days)  & $\mathcal{U}[ 11.234 , 11.236]$ &\Pbtwo[] \\
  \noalign{\smallskip}
    Transit epoch $T_0$ (BJD$_\mathrm{TDB}-$2\,450\,000)  & $\mathcal{U}[8939.77 , 8939.80]$ & \Tzerobtwo[]  \\ 
  \noalign{\smallskip}
    $e$ and $\omega$ polar parametrisation, $\sqrt{e} \cos \omega$ & $\mathcal{U}[-1,1]$  & \ecosbtwo \\
  \noalign{\smallskip}
    $e$ and $\omega$ polar parametrisation, $\sqrt{e} \sin \omega$ & $\mathcal{U}[-1,1]$  & \esinbtwo \\
  \noalign{\smallskip}
    Scaled planet radius $R_\mathrm{p}/R_{\star}$  &$\mathcal{U}[0.0,0.2]$ & \rrbtwo[]  \\
  \noalign{\smallskip}
    Impact parameter, $b$ & $\mathcal{U}[0,1]$ & \bbtwo[] \\
  \noalign{\smallskip}
    Stellar density $\rho_\star$ (${\rm g\,cm^{-3}}$) & $\mathcal{U}[0.1,1]$ & \denstrbtwo[]   \\
  \noalign{\smallskip}
    Doppler semi-amplitude variation $K$ (m s$^{-1}$) & $\mathcal{U}[0,500]$ & \kbtwo[] \\
    \multicolumn{3}{l}{\emph{ \bf Other sampled parameters}} \\
    Offset CHIRON (\kms) & $\mathcal{U}[ -28.51 , -27.12]$ & \CHIRONtwo[] \\
  \noalign{\smallskip}
    Offset FEROS (\kms) & $\mathcal{U}[ -27.08 , -25.68 ]$ & \FEROStwo[] \\
  \noalign{\smallskip}
    Offset TRES (\kms) & $\mathcal{U}[ -0.57 , 0.89]$ & \TREStwo[] \\
  \noalign{\smallskip}
    Linear trend (${\rm m\,s^{-1}\,d^{-1}}$) & $\mathcal{U}[ -1 , 1 ]$  & \ltrendtwo[] \\
  \noalign{\smallskip}
    Quadratic trend (${\rm m\,s^{-1}\,d^{-2}}$) & $\mathcal{U}[ -1 , 1 ]$  & \qtrendtwo[] \\
  \noalign{\smallskip}
    Jitter term $\sigma_{\rm CHIRON}$ (\ms) & $\mathcal{J}[1,100]$ & \jCHIRONtwo[] \\
  \noalign{\smallskip}
    Jitter term $\sigma_{\rm FEROS}$ (\ms) & $\mathcal{J}[1,100]$ & \jFEROStwo[] \\
  \noalign{\smallskip}
    Jitter term $\sigma_{\rm TRES}$ (\ms) & $\mathcal{J}[1,100]$ & \jTREStwo[] \\
  \noalign{\smallskip}
    TESS S23 limb-darkening coefficient $q_1$  &$\mathcal{U}[0,1]$ & \qoneSLtwo \\ 
  \noalign{\smallskip}
    TESS S23 limb-darkening coefficient $q_2$  &$\mathcal{U}[0,1]$ & \qtwoSLtwo \\ 
  \noalign{\smallskip}
    TESS S50 limb-darkening coefficient $q_1$  &$\mathcal{U}[0,1]$ & \qoneSCtwo \\ 
  \noalign{\smallskip}
    TESS S50 limb-darkening coefficient $q_2$  
    &$\mathcal{U}[0,1]$ & \qtwoSCtwo \\ 
  \noalign{\smallskip}
    LCO limb-darkening coefficient $q_1$  &$\mathcal{U}[0,1]$ & \qonegblctwo \\ 
  \noalign{\smallskip}
    LCO limb-darkening coefficient $q_2$  
    &$\mathcal{U}[0,1]$ & \qtwogblctwo \\
  \noalign{\smallskip}
    Jitter term $\sigma_{TESS,S23}$ (ppm) & $\mathcal{J}[1,100]$ & \jtrSLtwo[] \\
  \noalign{\smallskip}
    Jitter term $\sigma_{TESS,S50}$ (ppm) & $\mathcal{J}[1,100]$ & \jtrSCtwo[] \\
  \noalign{\smallskip}
    Jitter term $\sigma_{LCO}$ (ppm) & $\mathcal{J}[1,100]$ & \jtrgblctwo[] \\
    \multicolumn{3}{l}{\emph{ \bf \toitwob's derived parameters}} \\
    Planet mass $M_\mathrm{p}$ ($M_{\rm J}$) & $\cdots$ &  \mpbtwo[]     \\
  \noalign{\smallskip}
    Planet radius $R_\mathrm{p}$ ($R_{\rm J}$) & $\cdots$ &  \rpbtwo[]   \\
    Planet density $\rho_{\rm p}$ (g\,cm$^{-3}$) & $\cdots$ &  \denpbtwo[]    \\
  \noalign{\smallskip}
    Orbital eccentricity, $e$ & $\cdots$ & \ebtwo[] \\
  \noalign{\smallskip}
    Angle of periastron, $\omega$(deg) & $\cdots$ & \wbtwo[] \\
  \noalign{\smallskip}
    Scaled semi-major axis  $a/R_\star$ & $\cdots$ & \arbtwo   \\
  \noalign{\smallskip}
    Semi-major axis  $a$ (AU) & $\cdots$ &  \abtwo[]  \\
  \noalign{\smallskip}
    Time of periastron passage $T_p$ (BJD-2450000) & $\cdots$ & \Tzerobtwo[]  \\
  \noalign{\smallskip}
    Orbit inclination $i_\mathrm{p}$ ($^{\circ}$)  & $\cdots$ & \ibtwo[]    \\
  \noalign{\smallskip}
    Total transit duration $\tau_{14}$ (hours) & $\cdots$ & \ttotbtwo[]   \\
  \noalign{\smallskip}
    Planet surface gravity $g_{\rm p}$ (${\rm cm\,s^{-2}}$) & $\cdots$ & \grapbtwo[] \\
  \noalign{\smallskip}
    Equilibrium temperature  $T_\mathrm{eq}$ (K)\tablefootmark{c} & $\cdots$  &   \Teqbtwo[]  \\
  \noalign{\smallskip}
    Received irradiance ($F_\oplus$) & $\cdots$ & \insolationbtwo[]  \\
    \noalign{\smallskip}
    \hline
  \end{tabular}
\tablefoot{
\tablefoottext{a}{$\mathcal{U}[a,b]$ refers to an uniform prior between $a$ and $b$, $\mathcal{N}[a,b]$ to a Gaussian prior with mean $a$ and standard deviation $b$, and $\mathcal{J}[a,b]$ to the modified Jeffrey's prior as defined by \citet[eq.~16]{Gregory2005}.} 
\tablefoottext{b}{Inferred parameters and errors are defined as the median and 68.3\% credible interval of the posterior distribution.} 
\tablefoottext{c}{Assuming a zero albedo.}
}
\end{center}
\end{table*}

\section{Discussion} \label{sec:disc}

\subsection{The inferred formation mechanism of TOI-2420\,b and TOI-2485\,b}
\begin{figure*}
    \centering
    \includegraphics[trim = 0mm 0mm 0mm 80mm, clip, width=0.90\textwidth]{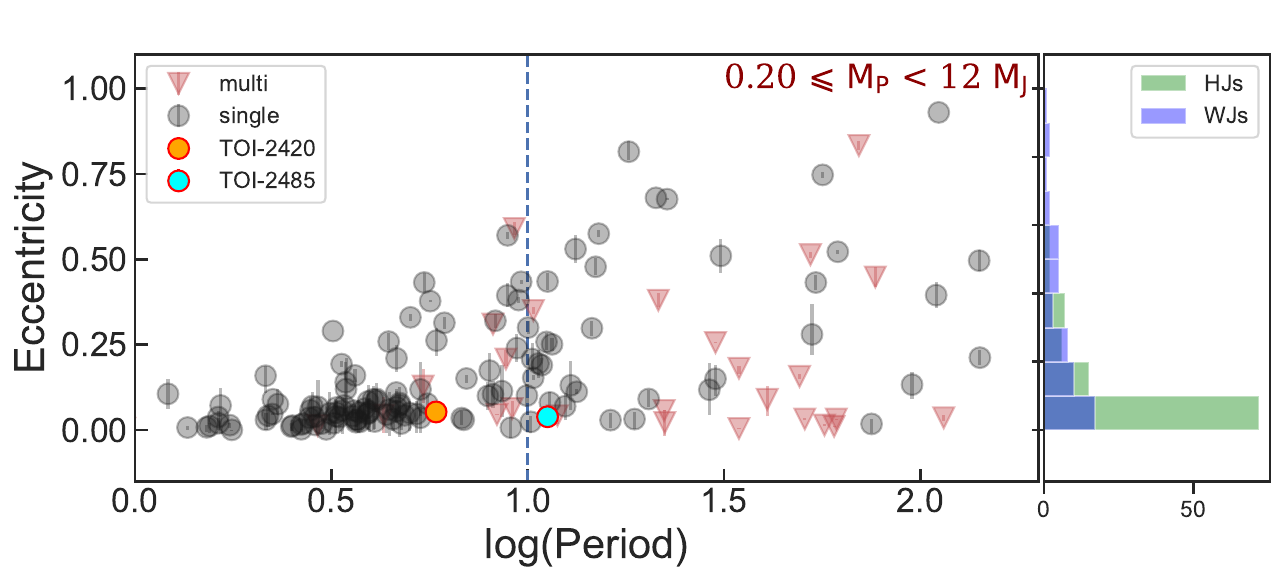}
    \caption{Eccentricity distribution as a function of the orbital period for Jupiter-sized planets (data taken as of UT 2024 March 13). The dashed blue line represents the 10-day boundary between HJs and WJs. The orange point represents TOI-2420\,b, while the cyan point represents TOI-2485\,b. The red triangles represent the planets in multi-planetary systems.}
    \label{fig:Jup_pop}
\end{figure*}

Considering the orbital period of 10 days as the boundary between HJs and WJs, TOI-2420\,b falls in the HJs category while TOI-2485\,b in the WJs category. Both planets are common outcomes in core accretion models including disk migration~\citep[e.g.,][]{IdaLin2013,Emsenhuber2021b,Schlecker2021a,Schlecker2021b}.
To put them in the context of the close-in giant planets population, from the NASA Exoplanet Archive\footnote{https://exoplanetarchive.ipac.caltech.edu/} we selected Jupiter-sized planets (mass between 0.20 and 12 M$_{J}$) with orbital periods shorter than 200 days, planetary masses with precision better than 20\% and eccentricities with precision smaller than 0.1. With these criteria, over a total of 5595 exoplanets (as of March 13th) we found 158 Jupiter-sized planets, of which 131 are in single systems and 27 in multi-planet systems. Fig. \ref{fig:Jup_pop} represents the eccentricity distribution of these two populations as function of the orbital period, showing that the majority of HJs orbiting within 3 days have circular orbits, while for increasing periods there is a wide variety of low and high eccentricities. This distribution challenges the evolution theories: disk migration cannot explain high eccentricities \citep[i.e.,][]{Bitsch2013, Petrovich2015,DuffellChiang2015}, while high-eccentricity tidal migration \citep{Wu2003} can explain the intermediate-high eccentricities of WJs with small pericenter distances, since tidal migration strongly depends on the distance from the star. The highly eccentric WJs might be the results of Kozai-Lidov oscillations or other secular oscillations caused by a third body \citep[i.e.,][]{Dong2014,PetrovichTremaine2016}. 

An important facet to consider in constraining these planets' dynamical histories is the tidal dissipation. Both planets' orbits are nearly circular (\eb[] for TOI-2420\,b and \ebtwo[] for TOI-2485\,b); could they have migrated by high-eccentricity migration and undergone tidal damping of eccentricity within the stars' main sequence lifetime? If not, we can exclude this migration method.

For current purposes we pursue a simple computation. We use the eccentricity tidal damping timescales presented in Eqn.~1 of \cite{DobbsDixon04}:
\begin{equation}
    \tau_{ep} \simeq 5 \bigg(\frac{Q'_P}{10^6}\bigg)\bigg(\frac{M_P}{M_J}\bigg)\bigg(\frac{M_{\star}}{M_{\odot}}\bigg)^{2/3}\bigg(\frac{P}{1 \mathrm{day}}\bigg)^{13/3}\bigg(\frac{R_P}{R_J}\bigg)^{-5} \mathrm{Myr}
\end{equation}
The tidal quality factor $Q'_P$ is significantly unknown, however, using a nominal value of $10^5$ predicts eccentricity damping timescales of 250 Myr and 32 Gyr for TOI-2420 and TOI-2485, respectively. This suggests that the former planet could have quickly tidally damped out any eccentricity for a range of tidal quality factors, while TOI-2485 would have required a low quality factor to damp out a high eccentricity within the age of the system.

We can also consider whether any initial orbital obliquity, as predicted by some high-eccentricity migration models, could have been damped out. \cite{Albrecht2012} presented a formula (Eqn.~2 of that work) for the obliquity tidal dissipation timescale of a convective-envelope star like TOI-2420 or TOI-2485, based upon the theory of \cite{Zahn1977} and calibrated using binary stars. This is:
\begin{equation}
    \frac{1}{\tau_{\mathrm{CE}}} = \frac{1}{10 \times 10^9 \mathrm{yr}} q^2 \bigg(\frac{a/R_{\star}}{40}\bigg)^{-6}
\end{equation}
where $\tau_{\mathrm{CE}}$ is the tidal damping timescale and $q$ is the planetary-to-stellar mass ratio. 

With the current parameters for both systems, the estimated obliquity damping timescales are $>10^{11}$ years, well in excess of the main sequence lifetimes in either system. Additionally, considering that both stars are slightly evolved and the current value of $a/R_{\star}$ is smaller than it was on the main sequence, these estimated tidal damping timescales may be \textit{underestimates}. 
Further information on their history can be obtained from the observations of the Rossiter-McLaughlin (RM) effect, as the long tidal obliquity timescales suggest that any initial orbital obliqity should be retained. The expected semi-amplitude of the RM is $\sim$5 \ms and $\sim$15 \ms for TOI-2420\,b and TOI-2485\,b, respectively. \citealt{toi2485RM} present the RM for TOI-2485\,b, showing that the system is well aligned, implying a quiet formation history.

\subsection{Evidence for a long-period companion to TOI-2485}

If we assume that the linear acceleration of TOI-2485 (-0.389 $\pm$ 0.009 m/s/days from the linear trend model) is caused by another orbiting body, we can place some constraints on the nature of that body and its orbit. Following equation 2 of \cite{2017MNRAS.464.2708S} (which is derived from equation 2 of \citealt{2002ApJ...571..519L}) we can place the following constraint on the mass and orbital separation of planet 'c':
\begin{equation}
\label{eqn:mass_c}
    \frac{M_\mathrm{c}}{a_\mathrm{c}^2} > 2.06 \, \mathrm{M_{Jup} \,au^{-2}}
\end{equation}

Furthermore, assuming an approximately circular orbit for 'c' implies an orbital period greater than twice our RV baseline, i.e. $P_\mathrm{c} > 1800$~d. This corresponds to $a_\mathrm{c} > 3$~au, and leads to (from Eqn.~\ref{eqn:mass_c}) a lower mass limit of 18~$\mathrm{M_{Jup}}$. For circular orbits, the third body would lie in the brown dwarf mass regime for $3 < a/\mathrm{au} < 6.3$. Alternative possibilities are a low-mass star orbiting further out, or a more massive object on a highly-eccentric orbit.
Using equation (4) of \cite{Jackson2021}, we are able to assess whether this putative third body is capable of inducing high-eccentricity tidal migration. A perturber at 3~au whose orbit is aligned with that of TOI-2485\,b, must have a mass greater than $7.3~\mathrm{M_{Jup}}$; in other words a 3~au, $18~\mathrm{M_{Jup}}$ object meets the minimum requirement for high-eccentricity tidal migration. 

RV surveys of WJs aimed at detecting long-term trends would be highly valuable in determining if perturber-coupled high eccentricity migration is the dominant mechanism. TOI-2485\,b, exhibiting a significant long-term trend, provides a promising connection to this hypothesis as suggested by \cite{Jackson2021}. 

\subsection{Planets orbiting evolved stars}

Both TOI-2420\,b and TOI-2485\,b orbit slightly evolved stars.
When a main sequence star exhausts the hydrogen in its core, the core contracts, while the temperature rises enough to ignite the fusion of hydrogen in a shell surrounding the core \citep[see, e.g.,][]{Lamers2017}. Prior to becoming a red giant, the evolving star undergoes a transition phase called subgiant phase, during which the star lies between the main sequence turn-off and the base of the red giant branch in the HR diagram \citep[see, e.g.,][]{Pinsonneault2023}. During the subgiant phase, the outer layers of the star expand, while the effective temperature decreases, profoundly affecting the evolution and the fate of the surrounding planetary system. The strong tidal interaction between a close-in planet and its expanding host star is expected to play a crucial role in shaping the structure of the inner region of a planetary system \citep{Villaver2009,Veras2016,Grunblatt2018,MacLeod2018}.

Based on radial velocity follow-up observations of about 500 bright (V \textless\ 8.5) sub-giant stars, the Lick, Keck, and California planet-search programs \citep{Johnsonetal2006,Satoetal2008,Peeketal2009,Luhnetal2019} found a paucity of hot Jupiters orbiting sub-giant stars with respect to main sequence stars. This result suggests that close-in planets might be engulfed by their evolving host star during the subgiant/giant phase \citep[e.g.,][]{VillaverLivio2009,Bowleretal2010,Villaveretal2014}.
\cite{Grunblattetal2019} carried out a search of transiting planets in a sample of nearly 2500 low-luminosity red giant branch stars observed by NASA K2 mission and found that short-period (P \textless\ 10 d) planets larger than Jupiter seem to be more common around evolved stars than main sequence stars. This would suggest that close-in planets larger than Jupiter can survive the subgiant phase, at least while their host stars have radii smaller than 5-6 R$_{\odot}$.

Only a few transiting planets with measured masses and radii, around sub-giant stars have been discovered so far from both ground- \citep[e.g.,][]{Lillo-Boxetal2016,Pepperetal2017,Grievesetal2021,Kabath2022,Smith2022} and space-based transit-search surveys \citep[e.g.,][]{2010ApJ...713L.126B,Roweetal2014,Ortizetal2015,Mortonetal2016,WangSonghuetal2019}. To increase our knowledge on the evolution of planetary systems during the post-main sequence phase of their host stars, it it crucial to increase the sample of well-characterized planets orbiting evolved stars.

\section{Conclusions}
This paper presents the discovery of two Jupiter-sized planets. TOI-2420\,b has been observed by TESS in Sectors 3 and 30, followed-up through LCO ground-based photometry and CHIRON, TULL and \textsc{Minerva}-Australis spectroscopy. We found TOI-2420\,b to have an orbital period of 5.8 days, a mass of 0.9 M$_{\rm J}$ and a radius of 1.3 R$_{\rm J}$. TOI-2485\,b has been observed by TESS during Sectors 23 and 50. We collected LCO photometry, and CHIRON, FEROS and TRES RVs data in order to constrain the orbital properties. TOI-2485\,b has an orbital period of 11.2 days, a mass of 2.4 M$_{\rm J}$ and a radius of 1.1 R$_{\rm J}$. 

The observed characteristics of the two planetary systems and the calculation of the tidal damping timescale indicates that the high-eccentricity migration (HEM) scenario cannot be ruled out for both systems, especially due to the large uncertainties in the tidal quality factor. Moreover, regarding TOI-2485\,b, the possible non-zero eccentricity and the evidence for a long period companion may support some HEM scenarios such as coplanar high-eccentricity migration \citep{Petrovich2015coplanar}.

\onecolumn
\appendix
\section{Tables of Radial velocity data.}

\begin{longtable}{llcc}
\caption{\label{tab:TOI2420_rv} Time series of TOI-2420 from CHIRON, Tull and Minerva-Australis data. We list the radial velocities and the corresponding uncertainties.}\\
\hline
 \hline
\noalign{\smallskip}
Dataset & JD-2450000  &      RV         &    $\sigma_{RV}$      \\
        &     & (m\, s$^{-1}$) &    (m\, s$^{-1}$)   \\
\hline
\noalign{\smallskip}
CHIRON      &  9399.914500   &  15823.7   &  21.5 \\
  & 9407.838120  &   15868.5  &   18.1  \\
  & 9409.872080  &   15932.4  &   27.0  \\
  & 9410.908300  &   15850.1  &   30.6  \\
  & 9412.840510  &   15808.3  &   18.7  \\
  & 9420.868650  &   15993.3  &   21.8  \\
  & 9421.839160  &   15990.6  &   27.5  \\
  & 9423.837000  &   15829.2  &   20.1  \\
  & 9425.911150  &   15970.1  &   22.6  \\
  & 9426.852700  &   16007.0  &   14.8  \\
  & 9427.840540  &   15861.5  &   11.5  \\
  & 9437.754440  &   15949.2  &   25.0  \\
  & 9440.820040  &   15813.7  &   20.8  \\
  & 9457.784050  &   15915.2  &   20.8  \\
  & 9459.777570  &   15885.8  &   13.5  \\
  & 9462.778500  &   15973.2  &   15.8  \\
  & 9463.731050  &   15828.4  &   49.1 \\ 
  & 9464.775950  &   15800.3  &   19.3 \\
\noalign{\smallskip}
\hline
\noalign{\smallskip}
Tull Coud{\'e}      &   9413.952528  &    10226.8   &    23.0    \\
  &    9454.914279  &    10233.8   &    17.7 \\
  &    9471.837927  &    10138.0   &    24.2 \\
  &    9472.859749  &    10265.6   &    22.3 \\
  &    9516.740156  &    10128.8   &    20.7 \\
  &    9528.706405  &    10131.7   &    23.6 \\
  &    9529.696991  &    10118.4   &    19.7 \\
  &    9541.641191  &    10176.8   &    22.3 \\
  &    9563.683901  &    10100.3   &    22.3 \\
  &    9592.605903  &    10177.9   &    24.4 \\
  &    9790.932952  &    10230.8   &    21.0 \\
  &    9791.928397  &    10179.6   &    21.3 \\
  &    9792.937453  &    10187.8   &    23.6 \\
  &    9797.948448  &    10133.0   &    20.2 \\
  &    9845.830299  &    10167.2   &    17.6 \\
   &   9846.815870  &    10259.8   &    22.5 \\
  &    9847.805267  &    10308.9   &    20.7 \\
  &    9848.825072  &    10234.3   &    23.0 \\
 &     9878.766502  &    10167.7   &    36.5 \\
 &     9906.660824  &    10346.0   &    24.3 \\
  &    9918.634488  &    10250.4   &    54.0 \\
  &    9926.634464  &    10146.6   &    29.5 \\
\noalign{\smallskip}
\hline
\noalign{\smallskip}
Minerva-Australis M3      &  9371.324892 &  16478.8  & 30.4   \\
 & 9378.305348  & 16471.4 &  39.6  \\
 & 9379.307379  & 16649.3 &  59.5  \\
 & 9381.297958  & 16524.1 &  42.3  \\
 & 9415.279225  & 16580.8 &  45.9  \\
 & 9421.263658  & 16757.6 &  23.0 \\
 & 9422.261202  & 16672.1 &  55.7  \\ 
 & 9425.181197  & 16492.9  & 30.1  \\
 & 9426.250063  & 16583.0 &  33.7  \\
 & 9426.298909  & 16616.1 &  48.2  \\
 & 9427.175827 &  16696.5 &  38.7  \\
 & 9430.235254  & 16436.6 &  42.9  \\
 & 9433.160739  & 16579.3 &  27.3  \\
 & 9434.217930  & 16491.3 &  45.7  \\
 & 9442.174683  & 16360.4 &  61.9  \\
 & 9443.204420  & 16422.2 &  37.0  \\
 & 9444.131234  & 16612.5 &  54.8  \\
 & 9448.121103 &  16316.4 &  46.1  \\
 & 9449.280400 &  16449.1 &  35.8  \\
 & 9453.107521 &  16447.0 &  41.3  \\
 & 9454.104662 &  16446.4 &  49.4  \\
 & 9467.258812 &  16626.2 &  14.9  \\
 & 9471.059332 &  16425.4 &  56.1  \\
 & 9473.058678 &  16559.2 &  43.3  \\
 & 9477.043370 &  16434.3 &  58.3  \\
 & 9479.038114 &  16709.2 &  36.6  \\
 & 9480.034754 &  16577.1 &  39.8  \\
 & 9481.031660 &  16432.4 &  39.5  \\
 & 9481.171649 &  16506.4 &  37.2  \\
 & 9482.062904 &  16410.1 &  64.2  \\
 & 9482.195567  & 16434.2 &  35.3  \\
 & 9483.026172 &  16281.9 &  59.5  \\
 & 9484.026661 &  16512.8 &  48.6  \\
 & 9486.018409 &  16623.8 &  51.1  \\
 & 9503.969131 &  16532.5 &  30.8  \\
 & 9504.976196 &  16515.4 &  68.7  \\
 & 9506.969859 &  16493.0 &  25.1  \\
 & 9514.937579 &  16645.8 &  99.0  \\
 & 9532.967716 &  16615.3 &  49.3  \\
 & 9534.936446 &  16328.3 &  61.7  \\
 & 9558.949592 &  16470.9 &  57.6  \\
 & 9559.949353 &  16501.0 &  62.3  \\
 & 9563.951425 &  16407.1 &  59.1  \\
\noalign{\smallskip}
\hline
\noalign{\smallskip}
Minerva-Australis M4      &  9371.324892 &  16422.1  & 39.1   \\
 & 9493.995930 &  16478.2  & 46.3   \\
 & 9494.148486 &  16466.1 &  59.9   \\
 & 9506.969859 &  16495.4 &  50.8   \\
 & 9511.946086 &  16423.2 &  63.0   \\
 & 9514.937579 &  16700.1 &  82.8   \\
 & 9522.341526 &  15921.3 &  33.0   \\
 & 9526.930799 &  16405.6 &  48.9   \\
 & 9530.993078 &  16527.6 &  40.0   \\
 & 9532.967716 &  16578.8  & 54.0   \\
 & 9534.031821 &  16412.7 &  57.5   \\
 & 9536.023893 &  16474.9 &  85.8   \\
 & 9558.949592 &  16437.9 &  40.0   \\
 & 9559.949353 &  16417.6 &  52.4   \\
 & 9563.951425 &  16425.2 &  18.7   \\
\noalign{\smallskip}
\hline
\noalign{\smallskip}
Minerva-Australis M5      &  9371.324892 &  16531.8  & 34.7   \\
 & 9378.305348  & 16537.2 &  38.9  \\
 & 9379.307379 &  16628.3 &  46.6   \\
 & 9381.297958 &  16615.5 &  59.8   \\
 & 9386.284127 &  16696.0 &  51.5   \\
\noalign{\smallskip}
\hline
\noalign{\smallskip}
Minerva-Australis M6      &  9371.324892 &  16544.2  & 50.5   \\
 & 9381.297958 &  16503.5 &  55.6   \\
 & 9415.279225 &  16638.0 &  59.3   \\
 & 9422.261202 &  16703.4 &  60.4   \\
 & 9425.181197 &  16584.2 &  27.4   \\
 & 9426.250063  & 16563.0 &  71.0   \\
 & 9426.298909  & 16698.2 &  72.4   \\
 & 9427.175827 &  16622.5 &  64.3   \\
 & 9430.235254 &  16446.1 &  59.8   \\
 & 9433.160739 &  16597.4 &  46.6   \\
 & 9434.217930 &  16440.7 &  60.3   \\
 & 9442.174683 &  16378.7 &  49.0   \\
 & 9443.204420 &  16470.6 &  65.8   \\
 & 9444.131234 &  16607.8 &  47.8   \\
 & 9448.121103 &  16548.4 &  43.6   \\
 & 9453.107521 &  16241.6 &  74.9   \\
 & 9454.104662 &  16458.2 &  48.9   \\
 & 9471.059332 &  16513.0 &  68.0   \\
 & 9473.058678 &  16557.1 &  61.7   \\
 & 9477.043370 &  16548.9 &  74.8   \\
 & 9479.038114 &  16798.9 &  85.5   \\
 & 9480.034754 &  16670.9 &  68.8   \\
 & 9481.031660 &  16478.4 &  48.8   \\
 & 9481.171649 &  16484.3 &  41.0   \\
 & 9482.062904 &  16420.5 &  36.8   \\
 & 9482.195567 &  16525.0 &  39.2   \\
 & 9483.026172 &  16457.8 &  65.5   \\
 & 9484.026661 &  16488.3 &  55.3   \\
 & 9486.018409 &  16709.9 &  67.1   \\
 & 9489.022530 &  16460.4 &  24.0   \\
 & 9490.007396 &  16614.0 &  46.4   \\
 & 9491.004888 &  16665.6 &  23.0   \\
 & 9492.002146 &  16632.1 &  43.6   \\
 & 9492.141531 &  16733.4 &  35.0   \\
\noalign{\smallskip}
\hline
\noalign{\smallskip}
\end{longtable}

\begin{longtable}{llcc}
\caption{\label{tab:TOI2485_rv} Time series of TOI-2485 from CHIRON, TRES and FEROS data. The radial velocities and corresponding uncertainties are listed.}\\
\hline
 \hline
\noalign{\smallskip}
Dataset & JD-2450000  &      RV         &    $\sigma_{RV}$      \\
        &     & (m\, s$^{-1}$) &    (m\, s$^{-1}$)   \\
\hline
\noalign{\smallskip}
CHIRON      &  9270.850340  & -27889.7  & 25.8  \\
 & 9273.828530  & -27620.7 &  24.1  \\
 & 9276.813900  & -27802.9 &  44.7  \\
 & 9279.778930  & -28011.6 &  17.7  \\
 & 9282.780180  & -27786.1 &  20.2  \\
 & 9305.714230  & -27737.6 &  14.3  \\
 & 9309.737680 &  -27711.2 &  27.1  \\
 & 9314.697910 &  -27919.0 &  20.7  \\
 & 9319.690760 &  -27636.6 &  18.7  \\
 & 9321.703320 &  -27810.6 &  22.6  \\
 & 9322.704030 &  -27910.3 &  12.2  \\
 & 9327.668190  & -27829.6 &  18.6  \\
 & 9336.639620 &  -27962.6 &  16.2  \\
 & 9345.643020 &  -27912.7 &  24.2  \\
\noalign{\smallskip}
\hline
\noalign{\smallskip}
TRES      &   9262.983173    &       387.8    &     30.4  \\
 & 9267.884753     &       -3.8   &      29.5  \\
 & 9268.976579    &       -67.2    &     16.9  \\
 & 9269.913158    &         4.7    &     15.9  \\
 & 9270.927043    &       103.5    &     17.0  \\
 & 9272.848224    &       280.8    &     34.8  \\
 & 9276.932610    &       122.1    &     19.6  \\
 & 9298.916023    &       177.5    &     23.3  \\
 & 9308.780538    &       304.7    &     15.0  \\
 & 9309.748096    &       264.3    &     18.6  \\
 & 9311.857518    &       -39.9    &     15.7  \\
\noalign{\smallskip}
\hline
\noalign{\smallskip}
FEROS      &  9267.863580 &  -26541.0 &  8.2 \\
 & 9270.811330 &  -26437.5 &  8.1  \\
 & 9273.812920 &  -26177.8 &  9.3  \\
 & 9278.783470 &  -26530.8 &  10.5  \\
 & 9279.794720 &  -26578.0  & - 7.7  \\
 & 9281.764600  & -26472.1  & 10.1  \\
 &  10030.778950 &  -26.8162 &  9.1   \\
 & 10031.817030  & -26.8521 &  9.6   \\
 & 10032.774580  & -26.8704 &  8.5   \\
 & 10047.721860 &  -26.5512 &  8.4   \\
 & 10048.753340 &  -26.4934 &  8.6   \\
 & 10067.676020 &  -26.8165 &  8.3   \\
 & 10125.490250 &  -26.6566 &  9.4   \\
 & 10127.462530 &  -26.5031 &  7.6   \\
 & 10128.513330 &  -26.4996 &  8.4   \\
\hline
\noalign{\smallskip}
\end{longtable}

\twocolumn

\begin{acknowledgements}
C.M.P. and M.F. gratefully acknowledge the support of the Swedish National Space Agency (DNR 65/19 and 177/19).
O.B. acknowledges that has received funding from the European Research Council (ERC) under the European Union’s Horizon 2020 research and innovation programme (Grant agreement No. 865624). 
G.N. thanks for the research funding from the Ministry of Education and Science programme the "Excellence Initiative - Research University" conducted at the Centre of Excellence in Astrophysics and Astrochemistry of the Nicolaus Copernicus University in Toru\'n, Poland.
P.K. acknowledges funding from LTT-20015 project.
D.G. gratefully acknowledges the financial support from the grant for internationalization (GAND\_GFI\_23\_01) provided by the University of Turin (Italy).
M.T.P. acknowledges support from the Fondecyt-ANID Post-doctoral fellowship 3210253 and from the ANID Project ASTRO21-0037.
A.J.\ and R.B.\ acknowledge support from ANID -- Millennium  Science  Initiative -- ICN12\_009.
R.B.\ acknowledges support from FONDECYT Project 1241963.
A.J.\ acknowledges support from FONDECYT project 1210718.

This work makes use of observations from the LCOGT network. Part of the LCOGT telescope time was granted by NOIRLab through the Mid-Scale Innovations Program (MSIP). MSIP is funded by NSF.
This research has made use of the Exoplanet Follow-up Observation Program (ExoFOP; DOI: 10.26134/ExoFOP5) website, which is operated by the California Institute of Technology, under contract with the National Aeronautics and Space Administration under the Exoplanet Exploration Program.
Funding for the TESS mission is provided by NASA's Science Mission Directorate. KAC and CNW acknowledge support from the TESS mission via subaward s3449 from MIT.

DRC and CAC acknowledge partial support from NASA Grant 18-2XRP18$\_$2-0007. This research has made use of the Exoplanet Follow-up Observation Program (ExoFOP; DOI: 10.26134/ExoFOP5) website, which is operated by the California Institute of Technology, under contract with the National Aeronautics and Space Administration under the Exoplanet Exploration Program.

\textsc{Minerva}-Australis is supported by Australian Research Council LIEF Grant LE160100001, Discovery Grants DP180100972 and DP220100365, Mount Cuba Astronomical Foundation, and institutional partners University of Southern Queensland, UNSW Sydney, MIT, Nanjing University, George Mason University, University of Louisville, University of California Riverside, University of Florida, and The University of Texas at Austin.

We respectfully acknowledge the traditional custodians of all lands throughout Australia, and recognise their continued cultural and spiritual connection to the land, waterways, cosmos, and community. We pay our deepest respects to all Elders, ancestors and descendants of the Giabal, Jarowair, and Kambuwal nations, upon whose lands the \textsc{Minerva}-Australis facility at Mt Kent is situated.

This research was carried out at the Jet Propulsion Laboratory, California Institute of Technology, under a contract with the National Aeronautics and Space Administration (80NM0018D0004).

The results reported herein benefitted from collaborations and/or information exchange within NASA’s Nexus for Exoplanet System Science (NExSS) research coordination network sponsored by NASA’s Science Mission Directorate under Agreement No. 80NSSC21K0593 for the program ``Alien Earths".

This work is partly supported by JSPS KAKENHI Grant Number JPJP24H00017 and JSPS Bilateral Program Number JPJSBP120249910.

 \end{acknowledgements}

\bibliographystyle{aa}
\bibliography{references}

\end{document}